\documentclass[12pt, draftclsnofoot, onecolumn]{IEEEtran}
\usepackage{times}
\usepackage[final]{graphicx}
\usepackage{textcomp}
\usepackage[reqno]{amsmath}
\usepackage{amsfonts}
\usepackage{times,amsmath,epsfig}
\usepackage{latexsym,amssymb}
\usepackage{cite}
\usepackage{psfrag}
\usepackage{stfloats}
\usepackage{amsmath}
\usepackage{amssymb}
\usepackage{array}
\usepackage[noend]{algorithmic}
\usepackage{algorithm}
\usepackage{bm}
\usepackage{color}
\usepackage{multicol}
\usepackage{multirow}
\usepackage{setspace}
\usepackage{tabularx}
\usepackage[T1]{fontenc}
\usepackage{subeqnarray}
\usepackage{cases}
\usepackage{makecell}
\newtheorem{thm}{\textbf{Theorem}}
\newtheorem{lem}{\textbf{Lemma}}
\newtheorem{prob}{\textbf{Problem}}
\newtheorem{remark}{\textbf{Remark}}
\usepackage{subfigure}
\newlength{\figwidth}
\newcommand{\PreserveBackslash}[1]{\let \temp =\\#1 \let \\ = \temp}
\newcolumntype{C}[1]{>{\PreserveBackslash\centering}p{#1}}
\newcolumntype{R}[1]{>{\PreserveBackslash\raggedleft}p{#1}}
\newcolumntype{L}[1]{>{\PreserveBackslash\raggedright}p{#1}}

\begin{document}

\title{Joint Task Offloading and Resource Allocation for IoT Edge Computing with Sequential Task Dependency}

\author{
Xuming An, Rongfei Fan, Han Hu, Ning Zhang, Saman Atapattu, and Theodoros A. Tsiftsis
\thanks
{
X. An, R. Fan, and H. Hu are with the School of Information and Electronics, Beijing Institute of Technology, Beijing 100081, P. R. China. (\{3120195381,fanrongfei,hhu\}@bit.edu.cn).
N. Zhang is with the Department of Electrical and Computer Engineering, University of Windsor, Windsor, ON, N9B 3P4, Canada. (ning.zhang@uwindsor.ca).
S. Atapattu is with the Department of Electrical and Electronic Engineering, The University of Melbourne, Parkville, VIC 3010, Australia. (saman.atapattu@unimelb.edu.au).
Theodoros A. Tsiftsis is with the Institute of Physical Internet and the School of Intelligent Systems Science and Engineering, Jinan University, Zhuhai 519070, China (e-mail: theodoros.tsiftsis@gmail.com).
}
}

\maketitle
\begin{abstract}
Incorporating mobile edge computing (MEC) in Internet of Things (IoT) enables resource-limited IoT devices to offload their computation tasks to a nearby edge server. In this paper, we investigate an IoT system assisted by the MEC technique with its computation task subjected to sequential task dependency, which is critical for video stream processing and other intelligent applications. To minimize energy consumption per IoT device while limiting task processing delay, task offloading strategy, communication resource, and computation resource are optimized jointly under both slow and fast fading channels. In slow fading channels, an optimization problem is formulated, {which is non-convex and involves one integer variable}. To solve this challenging problem,  we decompose it as a one-dimensional search of task offloading decision problem and a non-convex optimization problem with task offloading decision given. Through mathematical manipulations, the non-convex problem is transformed to be a convex one, which is shown to be solvable only with the simple Golden search method. In fast fading channels, optimal online policies depending on instant channel state are derived {even though they are entangled}. In addition, it is proved that the derived policy will converge to the offline policy when channel coherence time is low, which can help to save extra computation complexity. Numerical results verify the correctness of our analysis and the effectiveness of our proposed strategies over existing methods.
\end{abstract}


\section{Introduction} \label{s:intro}

{
The last decade has witnessed the fast-growing of the Internet of Things (IoT). This growth is even more expedited since the IoT meets artificial intelligence, through which patterns that matter can be learned from the massive data collected by IoT devices, and interested objects or anomalies can be spotted from instant data monitored by an IoT device \cite{8391395,zhou2019secure,guo2022constructing}.
On the other hand, the task of performing intelligent applications to detect interested objects or anomalies is usually of heavy burden, which is hard to be fully undertaken by any single IoT device, considering its limited computational capability, and the restriction on energy consumption.
Offloading the heavy computation task to the cloud center may be a solution for the IoT devices.
However, this will lead to a long time delay, which is especially undesirable when the running application is time-critical.
To overcome the above challenges, mobile edge computing (MEC) can offer a promising solution, which implements an edge server in the vicinity of the IoT devices and permits the IoT devices to offload their computation tasks to the edge server \cite{8016573,feng2020attribute}.
}

Regarding offloading the computation task \footnote{{When the task offloading is mentioned in this paper, it actually means to offload the input data of a computation task (which would be explained in Section \ref{s:model_task}). To avoid misleading, this course is uniformed to be {\it task offloading} throughout this paper.}}
from one IoT device to the edge server, {binary and partial offloading are mainly investigated. For binary offloading, the computation task is not separable. Hence it is completed locally at the IoT device or offloaded to and then computed at the edge server.
For partial offloading, it is assumed that} the computation task can be separated into two sub-tasks by any fraction, with one sub-task completed at the IoT device and the other one offloaded to and then computed in parallel at the edge server
\cite{8016573}.
{
Within the framework of either binary or partial offloading, the IoT device's energy consumption and task completion delay are the two most significant cost metrics in literature, which can be highly reduced by optimizing the task offloading decision, the allocation of communication and computation resources \cite{7879258,xu2021mcts,wang2016mobile,8488502,7842160,8387798,7762913}.
Specifically, task offloading decision determines how much data is left for local computations and offloading to the edge server, respectively \cite{7879258}; }
the communication resource involves not only IoT device's transmit power for task offloading \cite{wang2016mobile,8488502, mao2021computation}, but also the bandwidth \cite{7842160} or time slot duration \cite{8387798,7762913} for every IoT device when multiple devices are accessing the edge server; the computation resource primarily indicates edge server or IoT device's CPU frequency {\cite{huang2020result}} (which represents the computation capability) over time.
The general research goal is to minimize energy consumption per IoT device while promising the computation task completed within a tolerable delay, by jointly optimizing task offloading decision, the allocation of communication and computation resources.

{On the other hand, many intelligent applications, such as the deep neural network (DNN) driven applications including {digital forensics services\cite{ding2020deep}}, target recognition or anomaly detection {\cite{SIG-039}},
}
involve the type of computation task with multiple indivisible sub-tasks conforming to sequential dependency. To be specific, the whole computation task is comprised of multiple sub-tasks and every sub-task is executable only after completing its precedent sub-task, which is regarded as a {\it sequential task} in the following. Since a sequential task cannot be decomposed into two sub-tasks running in parallel, the strategies based on partial offloading in literature do not work anymore. Furthermore, although the strategies based on binary offloading in literature can help to generate a feasible solution for task offloading decision if we see the sequential task as a whole, energy consumption and time completion delay is not fully reduced since it can be further cut down when we choose to offload from some proper sub-task. Hence there exist model mismatch and performance degradation if the developed strategies based on partial offloading or binary offloading are utilized for our focused sequential task.

{
In terms of task offloading for sequential tasks,
given that the edge server has more powerful computation capability over the IoT device, it would be sensible to offload its computation task to the edge server as early as possible, i.e., to offload the sub-task at an early stage. However, due to the randomness of channel gain and the inhomogeneous input data amount associated with every sub-task, {a} long delay would be lead to in case of large input data amount and poor channel condition, no matter the offloaded sub-task is {at an early stage} or not. Hence it is not always good to offload the sub-task at an early stage, and it is interesting to study the best occasion to offload, which corresponds to a task offloading decision problem.}

{
Although there is some MEC literature working on the task offloading decision problem for the sequential task \cite{6849257,6846368}, which would be surveyed in Section \ref{s:related_work}, they do not explore other ways to reduce IoT device's energy consumption and task completion delay. It would be beneficial to further perform joint allocation of communication and computation resources since both of them can affect the above two types of cost metrics.
However, due to the entanglement of multiple types of variables which will lead to the non-convexity of the investigated problem and unpredictable random fluctuation of channel state, it is challenging to solve the focused problem directly. Additionally, this challenging problem cannot be answered well by the developed strategies based on partial or binary offloading due to the aforementioned model mismatch and performance degradation.
}

{In this work, we aim to solve the task offloading decision problem together with the allocation of communication and computation resources for the sequential tasks.}
Specifically, both slow and fast fading channels are considered.
Under these two considered channel models, task offloading strategy, communication resource, and/or computation resource are jointly optimized so as to minimize the energy consumption of the IoT device, while guaranteeing the task computation latency. {The main contributions are summarized as follows.}

{
\begin{itemize}
    \item {\textbf{System model:} We propose a framework of joint task offloading, communication and computation resource allocation for a special type of computation task in a MEC system, i.e., sequential task, which is particularly useful for describing intelligent applications such as DNN driven ones.
Fast fading and slow fading channels are investigated and associated optimization problems are formulated, which turn out to be a non-convex optimization problem involving one integer variable and a stochastic optimization problem with multiple online policies entangled in one constraint, respectively. Both of these two formulated problems are challenging to solve.
    }
    \item {\textbf{Joint task offloading and resource allocation under slow fading channel:}
    Global optimal solution of the non-convex optimization problem involving one integer variable is achieved through our two innovative efforts: 1) Divide the associated problem into two levels, with the lower level working on the non-convex part for any given integer variable and with the upper level dealing with the unique integer variable through one-dimensional exhaustive search; 2) Transform the non-convex problem in the lower level to be a convex one equivalently.
What is even further, simple Golden search is shown to be enough to find the optimal solution of the transformed convex problem by investigating its Karush-Kuhn-Tucker (KKT) condition.
}
    \item {\textbf{Joint task offloading and resource allocation under fast fading channel:}
The entangled multiple online policies are decomposed and the optimal online policies are found for the associated optimization problem through our two mathematical operations: 1) Divide the associated optimization problem into two levels, with the online offloading policy optimized freely from constraint in the upper level and the rest online policies optimized in the lower level for any given online offloading policy. 2) Explore some special properties of minimal expected energy consumption for task offloading to make the solving of the lower level problem possible.
Moreover, the dependency of cost function on instant channel state is proved to fade away when the channel coherence time approaches zero, which makes the replacement of online policy with offline policy at the price of ignorable performance loss but with much less computation complexity happen.}
    \item {\textbf{Numerical results:} Numerical results show that: 1) Our proposed methods can reduce energy consumption of IoT devices over the existing methods under both slow fading and fast fading channels; 2) The time consumption of our proposed methods for slow fading and fast fading channels are fully acceptable in the real application; 3) The disclosed properties of the minimal expected energy consumption for task offloading and founded convergence between online policy and offline policy as the channel coherence time approaches to zero are correct.}
\end{itemize}
}

The rest of the paper is organized as follows.
In Section \ref{s:related_work}, related literature is surveyed.
Section \ref{s:model} presents the system model and formulates the research problems under slow fading channel and fast fading channel.
In Section \ref{s:prob_solution}, the optimal solution for the formulated research problems under slow fading channel and fast fading channel are presented respectively, followed by the discussion when channel coherence time approaches zero. Numerical results are given in Section \ref{s:numerical_results} and concluding remarks are presented in Section \ref{s:conclusion}.

\section{Related works} \label{s:related_work}

In literature, plenty of works have studied binary offloading or partial offloading.
Due to the limit of space, we only survey the most representative works here, which includes the reference papers \cite{6574874,wang2016mobile,8387798,7762913,Add_NOMA_full_Ding,Add_NOMA_full_Peng,Add_NOMA_partial_Wu,8537962,Add_energy_harvesting,gu2020mobile}.
{
For binary offloading, \cite{6574874} proposes to take the action with less energy consumption between the options of fully local computing and fully offloading for a single-user MEC system.
\cite{wang2016mobile} investigates partial offloading problem together with the communication and computation resource allocation for a single-user MEC system.
\cite{8387798} and \cite{7762913} studies the joint resource allocation and partial offloading for a time division multiple access (TDMA) and orthogonal frequency division multiple access (OFDMA) MEC system, respectively. Since non-orthogonal multiple access (NOMA) can achieve higher spectrum efficiency over orthogonal multiple access (OMA), it is also taken into account for task offloading. With the setup of fully offloading, the benefit of NOMA for a MEC system is firstly analyzed and verified in \cite{Add_NOMA_full_Ding} and the resource allocation problem in a NOMA-aided MEC system is investigated in \cite{Add_NOMA_full_Peng}. For partial offloading, \cite{Add_NOMA_partial_Wu} studies task offloading and resource allocation jointly with decoding order fixed, while \cite{8537962} further optimizes the decoding order. Besides partial offloading, binary offloading, together with resource allocation, is also investigated in a NOMA-MEC system \cite{Add_NOMA_partial_Wu}.
In addition to the multiple access technique, the idea of energy-neutral is also brought into the MEC system, with every IoT device driven by the energy harvesting technique \cite{Add_energy_harvesting,gu2020mobile}.
For the energy harvesting MEC system,
partial offloading is investigated with the IoT device supported by randomly arrived environmental energy \cite{Add_energy_harvesting}, while both partial offloading and binary offloading are explored with the IoT device supported by wireless-power-transfer technique in \cite{gu2020mobile}.
}

Compared with the rich literature focusing on binary offloading or partial offloading, research works taking the sequential task model into account are very limited,  \cite{6849257,6846368,8854339}.
In \cite{6849257}, task completion delay is minimized for a single IoT device.
Not only the link rate but also the computation capability at the IoT device and edge server are supposed to be stable.
Hence only task offloading strategy is needed to be investigated.
The one-climb policy is disclosed to be optimal, in which the IoT device only offloads its computation task to the edge server once and only one segment of sub-tasks is computed at the edge server if ever.
With such a framework, \cite{6849257} figures out which segment of sub-tasks should be selected for the edge server
to achieve the minimal task completion delay.
In \cite{6846368}, a single IoT devices is considered and its expected energy consumption is minimized while limiting the expected or outage probability of task completion delay.
Three types of channel conditions are investigated, which are given as static channel, independently and identically distributed (i.i.d) block fading channel, and Markovian stochastic channel.
Similar to \cite{6849257}, only task offloading decision is explored and one-climb policy is proven to be optimal for all the investigated cases.
In \cite{8854339}, a MEC system with two IoT devices is considered under static channel conditions, both of which have a sequential task to compute.
Differently, one intermediate sub-task of the second IoT device requires the last sub-task's output of the first IoT device. A weighted sum of these two IoT devices' energy consumption and task completion delay is aimed to be minimized by optimizing not only the task offloading decision but also the CPU frequency and transmit power for offloading of these two IoT devices. With the task offloading decision given, the optimal solution of all the other variables is found. In terms of the task offloading decision, the one-climb policy is also proven to be optimal.

This research also investigates the MEC system under the sequential task model adopted in \cite{6849257} and \cite{6846368} for both slow fading channels and fast fading channels.
{
However, different from \cite{6849257} and \cite{6846368}, both of which solely focus on task offloading policy, we investigate the joint optimization of task offloading decision, communication resource (including IoT device's transmit power in every fading block) allocation, and/or computation resource allocation (including the allocated CPU for completing every sub-task).}
Although \cite{8854339} also investigates the joint design of task offloading decision and resource allocation,
{its working procedure involves inter-user task dependency, which is different from ours and the one in \cite{6849257} and \cite{6846368}. The research results published in \cite{8854339} cannot address the questions raised in this paper}.

\section{System Model and Problem Formulation}\label{s:model}
{Consider a MEC system with one edge server and multiple IoT devices. Every IoT device is allocated with one dedicated channel for task offloading and one dedicated CPU core at the edge server for computing. The bandwidth of the dedicated channel is $W$ and the dedicated CPU core at the edge server has a frequency of $f_e$
\footnote{Considering the incoming and outgoing of IoT devices in the coverage of the edge server, it would be computationally prohibitive if the edge server performs joint allocation of communication and computation resources among the connected IoT devices every once the set of connected IoT devices alters, which generally corresponds to a mixed-integer optimization problem. The dedicated allocation of spectrum and computation capability to every IoT device enable every IoT device to make its task offloading decision individually and separately, which can save the edge server from heavy and frequent optimization.}.
Without loss of generality and for the ease of presentation, we can study the MEC system by investigating one IoT device and the associated CPU core at the edge server, which is abbreviated to be one edge server for the ease of presentation, in the following.}
The IoT device has a computation task to complete. The task may be computationally intensive and has to be completed within time $T_{\text{th}}$. To complete the computation task in time and to save the IoT device's energy consumption, the IoT device can offload all or part of the computation task to the edge server on the allocated spectrum. In order to capture a good opportunity for task offloading, instant channel gain between the IoT device and edge server will be measured before the IoT device makes the decision to offload. After the edge server completes the computation, it will return computational results to the IoT device, which is often of small data size and requires little time to be transmited from the edge server to the IoT device \cite{8387798,7762913,8537962}.


\subsection{Computation Task Model} \label{s:model_task}
\begin{figure}
\begin{center}
  \includegraphics[scale=0.5]{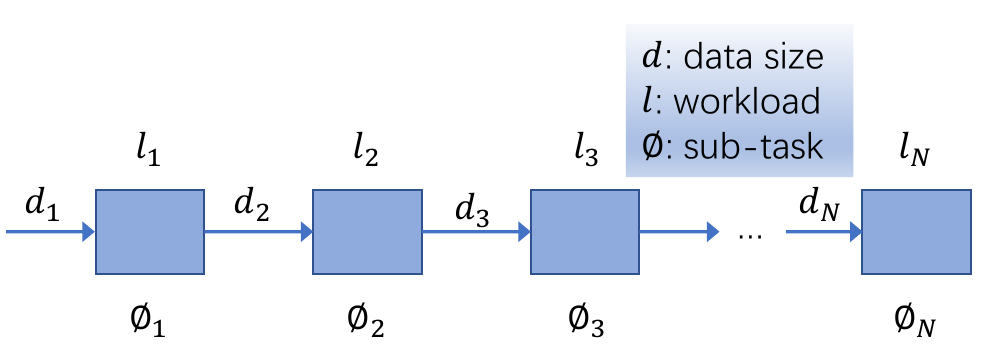}
\end{center}
\caption{{Sequentially dependent task model. There are $N$ consecutive sub-tasks in the mobile application
and all sub-tasks have to be executed sequentially. For $i \in \{2,...,N\}$, sub-task $\phi_i$'s input data, whose size is $d_i$, is also its precedent sub-task $\phi_{i-1}$'s output data.}}
\label{TM}
\end{figure}

For the computation task, a sequentially dependent task model is considered.
Specifically, the mobile application is assumed to be composed of $N$ sub-tasks, denoted as $\phi_1$, $\phi_2$, ..., $\phi_N$ respectively, the indices of which constitute the set $\mathcal{N} \triangleq \{1, 2, ..., N\}$. These $N$ sub-tasks have to be completed sequentially, i.e., one can only compute the sub-task of $\phi_i$ after completing the sub-task of $\phi_{i-1}$, for $i=2, 3, ..., N$.
Every sub-task can be characterized by two parameters. For sub-task $i\in \mathcal{N}$, the associated parameter pair is $(l_i, d_i)$, where $l_i$ indicates the amount of computation (which is in the unit of CPU cycles) and $d_i$ describes the amount of input data size (which is in the unit of nats for the ease of presentation). Note that with a sequentially dependent setup for the task model, the input data of sub-task $\phi_i$, which has a size of $d_i$, is also the output data of sub-task $\phi_{i-1}$.
{A diagram of the sequentially dependent task model is plotted in Fig. \ref{TM}}.
With the selection of the sequentially dependent task model, {the IoT device will first compute some sub-tasks at local and then offload the input data of the next sub-task to be computed to the edge server, at last the edge server will deal with the computing of the rest sub-tasks.
}

\subsection{Computation Model}
In terms of computation, it is composed of local computation both at the IoT device and edge server. Suppose the computation capability (which is also called CPU frequency) of the IoT device is $f_l \in [0, f_{\max}]$.
{Then the running power of the IoT device CPU is $k_0
f_l^3$ where $k_0$ is a physical parameter depending on the CPU's architecture according to \cite{BurdProcessor,6846368}.
Hence to complete a computation task with the computation amount being $l$, the time consumed is $l/f_l$ and the consumed energy is $k_0 f_l^2 l$.}
{
To save energy consumption of {the} IoT device considering that it is energy sensitive and its running power is largely determined by CPU frequency,
$f_l$ is assumed to be adjustable over disjoint sub-tasks. }
For sub-task $i \in \mathcal{N}$, suppose the associated CPU frequency is $f_i$, then to complete a sub-task, say $\phi_i$, the associated time consumption would be

\begin{equation} \label{e:time_local_comp}
t_m (i) =  \frac{l_i}{f_i}, \forall i \in \mathcal{N},
\end{equation}

and the associated energy consumption would be

\begin{equation} \label{e:energy_local_comp}
e_m (i) = k_0 \times l_i \times f_i^2, \forall i \in \mathcal{N}.
\end{equation}

The computation capability of the edge server, i.e., the CPU frequency, is given as $f_e$ where $f_e > f_{\max}$. According to the previous discussion, to complete a sub-task, say $\phi_j$, the time consumption would be

\begin{equation} \label{e:time_edge_comp}
t_e (j) = \frac{l_j}{f_e}, \forall j \in \mathcal{N}.
\end{equation}

\subsection{Communication Model} \label{s:communication_model}
The communication activity in the system encompasses uploading the data of some sub-task from the IoT device to the edge server, and downloading the data of computational results from the edge server to the IoT device.
In terms of data downloading, the data {size} of computational results is always small. In addition, the edge server has {a} strong power supply and can transmit information with {a} high data rate. Hence the time consumption for data downloading is omitted. This assumption is broadly adopted \cite{7762913,gu2020mobile}.

{In terms of task offloading, if the IoT device decides to stop local computing by sub-task $\phi_{n-1}$, then it only needs to offload the input data for computing sub-task $\phi_n$ to the edge server, which has an amount of $d_n$ nats.
It is worth mentioning that the input data of any subsequent sub-task, including $\phi_{n+1}, \phi_{n+2}, ..., \phi_N$, is also the output data of its precedent sub-task and can be generated after completing its precedent sub-task at the edge server, considering the sequential dependency between adjacent sub-tasks. Hence the IoT device does not need to offload the input data of any sub-task among $\phi_{n+1}, ..., \phi_N$ to the edge server.
}

In the process of offloading these $d_n$ nats, {the corresponding {\it normalized channel} gain from the IoT device to the edge server is denoted as ${\bm{h}_n}$ for $n\in \mathcal{N}$ \footnote{{The $\bm{h}_n$ may be a scalar or vector, as explained at the end of this paragraph. To be uniform, we write it in the form of bold font.}}.
The normalized channel gain represents the received SNR at the edge server when the transmit power is 1. If the noise power at the edge server is 1, the normalized channel gain will coincide with the real channel gain. For ease of discussion in the following, we abbreviate every normalized channel gain as channel gain.}
When the timespan of task offloading is less than the channel coherence time, which is denoted as $\tau$, the {$\bm{h}_n$} would be a single random variable. Otherwise, the {$\bm{h}_n$} should be split into multiple independently distributed fading blocks, which can be denoted as {$\bm{h}_n = \left(h_{n,1}, h_{n,2}, h_{n,M(n)}\right)^T$}. Detailed discussion on this point will be offered in the rest part of this subsection.
In addition, {the $\bm{h}_n$ no matter in case of scalar or vector} is supposed to be i.i.d.

By comparing the time-scale of task offloading and sub-task computation at local with the channel coherence time $\tau$, two categories of channel models can be expected.


\subsubsection{Slow fading channel} \label{s:communication_model_one}
In this type of channel model, channel coherence time would be longer than $T_{\text{th}}$.
In this case, {$\bm{h}_i$} does not change with $i$ for $i \in \mathcal{N}$ in the whole process of completing the computation task, no matter the sub-task is being processed locally, offloaded, or processed at the edge server. Denote the unchanged {$\bm{h}_i$} for $i\in \mathcal{N}$ as $h$, then the IoT device only needs to perform one measurement of $h$ before it starts to execute the computation task.

With such an assumption, denote the transmit power of the IoT device as $P_t$ when it is executing task offloading. {Considering the fact that $h$ is the normalized channel gain, the received SNR at the edge server can be written as $P_t h$ and the data transmission rate between the IoT device and edge server can be expressed as $r= W \ln \left(1 + P_t h\right)$}. In order to offload $d_n$ nats when the IoT device stops local computing by sub-task $\phi_{n-1}$, the time consumption is

\begin{equation} \label{e:time_offload_one_fading}
T_{\text{off}}^{\text{I}}(n) = \frac{d_n}{W \ln \left(1 + P_t h\right)}, \forall n \in \mathcal{N}
\end{equation}

and the energy consumption can be written as

\begin{equation} \label{e:energy_offload_one_fading}
E_{\text{off}}^{\text{I}}(n) = \frac{d_n P_t}{W \ln \left(1 + P_t h\right)}, \forall n \in \mathcal{N}.
\end{equation}

\subsubsection{Fast fading channel} \label{s:communication_model_multiple}
In this type of channel model, channel coherence time $\tau$ is much shorter {than not only the time scale of computing any sub-task but also the time scale of offloading the input data of any sub-task}.
Therefore, for $n \in \mathcal{N}$, the task offloading time window, whose duration is {supposed to be} $T(n)$, may contain multiple (say $M(n)$) fading blocks, which are indexed as $(n, 1)$, $(n, 2)$, ..., $(n, M(n))$, respectively.
In this case, {$\bm{h}_n$} is a vector rather than a scalar, which is composed of independently distributed $h_{n,1}$, $h_{n,2}$, ..., $h_{n, M(n)}$, for $n \in \mathcal{N}$. In addition, we suppose the random variables $h_{n,1}$, $h_{n,2}$, ..., $h_{n, M(n)}$ are also subject to a common distribution function $p(\cdot)$ with a sample space $\mathcal{H}$.

Within $T(n)$, there are $M(n)$ fading blocks, {which can be calculated as $M(n) = \lceil T(n)/\tau \rceil$}. Define $t_{n,m}$ as the time duration of $m$th fading block for $n \in \mathcal{N}$, $m \in \{1, 2, ..., M(n)\}$, then there is
\begin{equation} \nonumber{
{t_{n,m}} = \begin{cases}
\tau , & m < M(n), \\
   \tilde{t}_n , & m = M(n),
\end{cases}
}
\end{equation}
where $\tilde{t}_n\in (0, \tau]$ characterizes the time duration of the ending fading block for task offloading, which may be less than $\tau$ since $T(n)$ may not be exactly integer multiple of channel coherence time $\tau$.
Hence $T(n)$ satisfies
$T(n) = (M(n)-1)\tau + \tilde{t}_n, \forall n \in \mathcal{N}$.

When the IoT device decides to offload the input data for computing sub-task $\phi_n$, it will experience $M(n)$ fading blocks.
In one fading block whose time duration is $t_{n,m}$ and channel gain is $h_{n,m}$, then to offload $d$ nats, according to Shannon capacity formula,  the transmit power can be written as ${\left(e^{\frac{d}{W t_{n,m}}} - 1 \right)}/{h_{n,m}}$, and the associated energy consumption is
%
$e(d, h_{n,m}, t_{n,m}) = {\left(e^{\frac{d}{Wt_{n,m}}} - 1\right)t_{n,m}}/{h_{n,m}}$.
%
Define $Q_{n,m}(d, h; M(n),\tilde{t}_n)$ as the minimum amount of energy that the IoT device will consume for task offloading over the rest $(M(n) - m + 1)$ fading blocks under the following four conditions:
1) the channel gain at the beginning of fading block $(n, m)$, i.e., $h_{n,m}$, is measured to be $h$;
2) there are $d$ data nats to offload;
3) the total number of fading blocks for the task offloading is $M(n)$;
4) the last fading block, i.e, fading block $(n,M(n))$, has a time duration of $\tilde{t}_n$.
Define $Q_{n,m}(d; M(n), \tilde{t}_n)$ as the minimum amount of energy that the IoT device will consume for the task offloading over the rest $(M(n) - m + 1)$ fading blocks when the IoT device has no knowledge of channel gain $h_{n,m}$ at the beginning of fading block $(n, m)$ and the above mentioned conditions 2, 3, and 4 are satisfied.

With such a definition, when the IoT device finishes computing sub-task $\phi_{n-1}$, it will first measure the channel gain $h_{n,1}$, then evaluate the function  $Q_{n,1}(d_n, h_{n,1}; M(n), \tilde{t}_n)$, and finally make the decision on whether to offload the data for computing sub-task $\phi_n$ based on the evaluated $Q_{n,1}(d_n, h_{n,1}; M(n), \tilde{t}_n)$. How to make the decision according to $Q_{n,1}(d_n, h_{n,1}; M(n), \tilde{t}_n)$ will be discussed in Section \ref{OS3}.
According to the discussion in \cite{gu2020mobile}, the functions of $Q_{n,m}(d, h; M(n), \tilde{t}_n)$ and $Q_{n,m}(d; M(n), \tilde{t}_n)$ for $m\in \{1, 2,..., M(n)\}$ and $n\in \mathcal{N}$ can be calculated iteratively by following (\ref{e:Q_int}) and (\ref{e:Q_iterative}).

\begin{figure*}
\begin{equation} \label{e:Q_int}
	 {Q_{n,m}}\left(d; M(n), \tilde{t}_n\right)
	=  \int_{0}^{\infty} {{Q_{n,m}}\left(d,{h_{n,m}; M(n), \tilde{t}_n}
	\right)p\left( h_{n,m}\right)d{h_{n,m}}} , \forall m \in \left\{ {1,...,M(n)} \right\}, \forall n \in \mathcal{N},
\end{equation}
\end{figure*}


\begin{figure*}
\begin{equation} \label{e:Q_iterative}
	\begin{split}
	&  Q_{n,m}(d,{h_{n,m}}; M(n), \tilde{t}_n) \\
	 =  & \mathop {\min }\limits_{0 \leq {d_{n,m}} \leq d} \bigg( e(
	{{d_{n,m}},{h_{n,m}}}, t_{n,m}) +
	 \int_{0}^{\infty}  {{Q_{n,m+1}}\left( {d - d_{n,m},{h_{n,m+1}}; M(n), \tilde{t}_n} \right)
		p\left( {{h_{n, m+1}}} \right)d{h_{n,m+1}}} \bigg), \\
 = & \mathop {\min }\limits_{0 \leq {d_{n,m}} \leq d} \bigg( e(
	{{d_{n,m}},{h_{n,m}}}, t_{n,m}) +
	 {{Q_{n,m+1}}\left( d - d_{n,m}; M(n), \tilde{t}_n \right)} \bigg),
 \forall m \in \{1, 2,...,M(n)-1\}, \forall n \in \mathcal{N}.
	\end{split}
\end{equation}
\end{figure*}


The $d_{n,m}$ in (\ref{e:Q_iterative}) represents the amount of data to be offloaded within fading block $(n,m)$,
and

\begin{equation} \label{e:Q_init}
\begin{split}
	&{Q_{n,M(n)}}\left({d,h_{n,M(n)}}; M(n), \tilde{t}_n\right) \\
	=& e\left( {d,{h_{n,M(n)}},t_{n,M(n)}} \right) \\
	=& \frac{\left({e^{\frac{d}{\tilde{t}_nW}}} - 1\right) \tilde{t}_n}{{h_{n,M(n)}}}, \forall n \in \mathcal{N}.
\end{split}
\end{equation}

\subsection{Problem Formulation}\label{s:p}

With the two channel models introduced in Section \ref{s:communication_model}, our target is to minimize the total energy consumption of the IoT device while guaranteeing the computation task is completed within time $T_{\text{th}}$. Specifically, the associated optimization problems are given as follows.

\subsubsection{Slow fading channel} \label{s:prob_for_I}

In this case, the total energy consumption of the IoT device is composed of two parts: 1) The energy consumption for local computing; 2) The energy consumption for task offloading.
Hence combining the expressions in (\ref{e:energy_local_comp}) and (\ref{e:energy_offload_one_fading}), the total energy consumption if the IoT device decides to offload data exactly after completing sub-task $\phi_{n-1}$ can be written as

\begin{equation} \label{e:energy_consum_one_slot}
E_1(n) = \frac{{{d_n}
 {P_t}}}{{W \ln \left( {1 + {P_t} h} \right)}} +
\sum\limits_{i = 1}^{n{\rm{ - 1}}} {{l_i}  f_i^2 k_0}, \forall n \in \mathcal{N},
\end{equation}
where the second term of the right-hand side of (\ref{e:energy_consum_one_slot}) represents the energy consumption of the IoT device for computing sub-tasks $\phi_1$, $\phi_2$, ..., $\phi_{n-1}$ at local.

For the time consumption to complete the mobile application, if the IoT device decides to offload data exactly after completing sub-task $\phi_{n-1}$, by recalling (\ref{e:time_local_comp}), (\ref{e:time_edge_comp}), and (\ref{e:time_offload_one_fading}), the total time consumption, denoted as $T_1(n)$, can be written as

\begin{equation} \label{e:time_consum_one_slot}
T_1(n) = \sum\limits_{i = 1}^{n{\rm{ -
	1}}}{\frac{l_i}{f_i}} + \frac{d_n}{W \ln \left({1+P_t h}\right)} + \sum\limits_{i = n}^N {\frac{l_i}{f_e}}, \forall n \in \mathcal{N}.
\end{equation}

Then we need to select the stopping time $n\in \mathcal{N}$ and optimize $P_t$ and  $f_i$ for $i\in \{1,2,...,n-1\}$, so as to minimize the energy consumption in (\ref{e:energy_consum_one_slot}) while requiring the $T_1(n)$ to be no larger than $T_{\text{th}}$. Specifically, to minimize the total energy consumption of the IoT device, the following optimization problem needs to be solved
\footnote{{Since the running time for working out the optimal solution of Problem \ref{s:p:constant_gain} by the IoT device is ignorable compared with the time scale of $T_{\text{th}}$, which will be demonstrated in Section \ref{s:num_slow}, it is not necessary to take into account this part of time consumption in (\ref{p1_1}). Even if it has to be accounted for, the $T_{\text{th}}$ in constraint (\ref{p1_1}) can be replaced with $\left(T_{\text{th}}  - \Delta_c\right)$, where $\Delta_c$ represents the running time for solving the optimization problem. A similar discussion also follows for constraint (\ref{e:s:p:multiple_deadline}) in Problem  \ref{s:p:multiple}}.}
\begin{prob}\label{s:p:constant_gain}

\begin{subequations}
\begin{align}
	\mathop{\min} \limits_{P_t,n,\{f_i\}} & \quad \frac{d_n
 P_t}{W \ln \left( {1 + {P_t} h} \right)} +
\sum\limits_{i = 1}^{n{\rm{ - 1}}} {{l_i}  f_i^2 k_0} \nonumber \\
	\text{s.t.} & \quad
	\sum\limits_{i = 1}^{n - 1} \frac{l_i}{f_i} + \frac{d_n}{W \ln \left({1+P_t h}\right)} + \sum\limits_{i = n}^N {\frac{l_i}{f_e}} \le T_{\text{th}} , \label{p1_1} \\
	& \quad 0 \leq {f_i} \le {f_{\max}} , \forall i \in \{1,2,...,n-1\},  \\
    & \quad P_t \geq 0, \\
    & \quad n \in \mathcal{N}.
\end{align}
\end{subequations}

\end{prob}

\subsubsection{Fast fading channel} \label{s:prob_for_III}
In this case, \\$h_{1,1}, ..., h_{1,M(1)}, h_{2,1}, ..., h_{2,M(2)}, ..., h_{N,1}, ..., h_{N,M(N)}$ are all i.i.d random variables. The IoT device cannot determine when to offload data in advance, due to the randomness of channel gains in forthcoming fading blocks. Therefore, an online policy is required to be designed.

When the IoT device finishes the computing of sub-task $\phi_{n-1}$ at local, it can measure the channel gain $h_{n,1}$ and then makes the decision on whether to offload the input data of sub-task $\phi_n$ right now.
Hence before the IoT device determines to offload the computing data of sub-task $\phi_n$, it has the knowledge of {$\bm{h}_1, \bm{h}_2, ..., \bm{h}_{n-1}, h_{n,1}$}.
In this case, the stopping rule $n$ can be expressed as {$n\left(\bm{h}_1,...,\bm{h}_{n-1}, h_{n,1}\right)$ or $n\left(\bm{h}_1 \rightarrow h_{n,1}\right)$} for brevity.
Without confusion and for ease of presentation, we will interchange the use of $n$ and {$n\left(\bm{h}_1 \rightarrow h_{n,1}\right)$} in the following discussion for this scenario.
In terms of $f_i$ for $i\in \{1, 2, ..., n-1\}$, from the perspective of the IoT device, it is unclear when it will switch to task offloading, not to mention how to dynamically adjust $f_i$ over $i\in \{1,2,...,n-1\}$. To be equal with every sub-task, the IoT device will fix $f_i$ at $f_l \in [0, f_{\max}]$ for $i\in \{1, 2, ..., n-1\}$.

In this scenario, the time delay for completing the whole mobile application can be written as
\begin{equation}
T_3(n, M(n), \tilde{t}_n) = \sum\limits_{i = 1}^{n -
	1}{\frac{l_i}{f_l}} + \left(M(n)-1\right)\tau + \tilde{t}_n + \sum\limits_{i = n}^N {\frac{l_i}{f_e}}, \forall n \in \mathcal{N}.
\end{equation}
Then to meet the deadline requirement, there should be
%
$T_3(n, M(n), \tilde{t}_n) \leq T_{\text{th}}$.
%

The incurred energy consumption when the IoT device decides to stop local computing by sub-task $\phi_{n-1}$ can be written as

\begin{equation}
\begin{array}{ll}
  & E_3(n, M(n), \tilde{t}_n) \\
= & Q_{n,1}(d_n, h_{n,1}; M(n), \tilde{t}_n) +
\sum\limits_{i = 1}^{n- 1} {{l_i}  f_l^2 k_0}, \forall n \in \mathcal{N},
\end{array}
\end{equation}
then to minimize the expected total energy consumption of the IoT device, the associated optimization problem can be formulated as
\begin{prob}\label{s:p:multiple}

\begin{subequations}
	\begin{align}
		\mathop {\min }\limits_{\substack{{n({\bm{h}}_1\rightarrow h_{n,1})}, M(n), \tilde{t}_n}} \quad & \mathbb{E}_{{\bm{h}_1, ..., \bm{h}_N}} \left\{E_3(n, M(n), \tilde{t}_n)\right\} \nonumber \\
		\text{s.t.} \quad & {n({\bm{h}}_1\rightarrow h_{n,1})} \in \mathcal{N}, \\
		    &	T_3(n, M(n), \tilde{t}_n) \leq T_{\text{th}}, \label{e:s:p:multiple_deadline}\\
		           & M(n) \in \mathcal{I},\\
                   & 0 < \tilde{t}_n  \leq \tau,
	\end{align}
\end{subequations}

\end{prob}
where $\mathcal{I}$ is the integer set.

In Problem \ref{s:p:multiple}, it should be noticed that not only {$n(\bm{h}_1\rightarrow h_{n,M(n)})$} but also $M(n)$ and $\tilde{t}_n$ are all online policies depending on {$\bm{h}_1, \bm{h}_2, ..., \bm{h}_{n-1}, h_{n,1}$}, and the objective function represents the expected energy consumption $E_3(n, M(n), \tilde{t}_n)$ over the possible realizations of {$\bm{h}_1, \bm{h}_2, ..., \bm{h}_N$}.
{
Note that although $n(\bm{h}_1 \rightarrow h_{n,1})$, $M(n)$, and $\tilde{t}_n$ are all random, they are not necessarily independent. Hence a restriction like (\ref{e:s:p:multiple_deadline}) can be imposed on these three random variables, which makes them to be correlated.
}

\section{Optimal Solution} \label{s:prob_solution}

\subsection{Optimal Solution of the Task Offloading and Resource Allocation Problem in Slow Fading Channels} \label{OS1}

In this subsection, the problem of task offloading and resource allocation in slow fading channels, which corresponds to Problem \ref{s:p:constant_gain}, would be solved.
Problem \ref{s:p:constant_gain} involves the optimizing of variables $n\in \mathcal{N}$, $P_t$, and $f_i$ for $i\in \{1, 2, ..., n-1\}$, which contains both integer and continuous variables, and thus cannot be solved directly.
To find the optimal solution of Problem \ref{s:p:constant_gain}, we decompose it into two levels. In the {lower} level, $n$ is fixed while the rest variables $P_t$ and $f_i$ for $i\in \{1, 2, ..., n-1\}$ are optimized. The associated optimization problem can be written as
\begin{prob}\label{p:I_lower}

\begin{subequations}
\begin{align}
	E_{I}(n) \triangleq
	\mathop{\min} \limits_{P_t,\{f_i\}} \quad & \frac{d_n
 P_t}{W \ln \left( {1 + {P_t} h} \right)} +
\sum\limits_{i = 1}^{n- 1} {k_0 l_i  f_i^2} \nonumber \\
	\text{s.t.} &
	\sum\limits_{i = 1}^{n - 1} \frac{l_i}{f_i} + \frac{d_n}{W \ln \left({1+P_t h}\right)} + \sum\limits_{i = n}^N {\frac{l_i}{f_e}} \le T_{\text{th}} , \label{e:I_lower_active_inequality} \\
	& 0 \leq {f_i} \le {f_{\max}} , \forall i \in \{1,2,...,n-1\}, \\
    & P_t \geq 0.
\end{align}
\end{subequations}

\end{prob}
In the upper level, optimal $n$ should be found so as to solve the following optimization problem
\begin{prob} \label{p:I_upper}

\begin{equation}
\mathop{\min} \limits_{n \in \mathcal{N}} \quad E_{I}(n)
\end{equation}

\end{prob}

It can be checked that the upper level problem, Problem \ref{p:I_upper}, is equivalent with the original optimization problem, Problem \ref{s:p:constant_gain}. In the following, the lower level problem, Problem \ref{p:I_lower} will be solved first.
{Inspecting the function $\frac{d_n P_t}{W \log \left(1+ P_t h\right)}$ in the objective function of Problem \ref{p:I_lower},
it can be found that its second-order derivative with $P_t$ is always non-positive, which implies the concavity of the function $\frac{d_n P_t}{W \log \left(1+ P_t h\right)}$.
Thus Problem \ref{p:I_lower} is a non-convex optimization problem, considering that Problem \ref{p:I_lower} includes minimizing a concave function.
}
Problem \ref{p:I_lower} is a non-convex optimization problem, since the function $\frac{d_n P_t}{W \log \left(1+ P_t h\right)}$ in the objective function of Problem \ref{p:I_lower} is non-convex with $P_t$. To simplify the solving of Problem \ref{p:I_lower}, we define an assistant variable $\tau_t = \frac{d_t}{W\log\left(1+P_t h\right)}$, which represents the time consumption for offloading the input data of sub-task $\phi_n$. By expressing $P_t$ with $\tau_t$, i.e., $P_t = \left(e^{\frac{d_t}{W \tau_t}} - 1\right)\frac{1}{h}$, Problem \ref{p:I_lower} turns to be the following optimization problem
\begin{prob}\label{p:I_lower_convex}

\begin{subequations}
\begin{align}
	E_{I}(n) \triangleq
	\mathop{\min} \limits_{\tau_t,\{f_i\}} & \quad \tau_t \left(e^{\frac{d_n}{W \tau_t}} - 1\right) \frac{1}{h} +
\sum\limits_{i = 1}^{n- 1} {k_0 l_i  f_i^2} \nonumber \\
	\text{s.t.} & \quad
	\sum\limits_{i = 1}^{n - 1} \frac{l_i}{f_i} + \tau_t + \sum\limits_{i = n}^N {\frac{l_i}{f_e}} \le T_{\text{th}} , \label{e:I_lower_convex_active_inequality} \\
	& \quad 0 \leq {f_i} \le {f_{\max}} , \forall i \in \{1,2,...,n-1\}, \label{e:I_lower_convex_f_interval} \\
    & \quad \tau_t \geq 0.
\end{align}
\end{subequations}

\end{prob}
For Problem \ref{p:I_lower_convex}, it can be checked that the function $\tau_t \left(e^{\frac{d_n}{W \tau_t}} - 1\right) \frac{1}{h}$ is convex with $\tau_t$ since its second-order derivative with $\tau_t$ is larger than 0. In addition, both the function ${k_0 l_i  f_i^2} $ and the function $\frac{l_i}{f_i} $ are convex with $f_i$ for $i \in \{1, 2, ..., n-1\}$. Hence Problem \ref{p:I_lower_convex} is a convex optimization problem, which can be solved by traditional methods, such as the elliptical method or interior method \cite{boyd_vandenberghe_2004}.

To further simplify the solving of Problem \ref{p:I_lower_convex}, we target to design a simple yet optimal solution for Problem \ref{p:I_lower_convex}.
To explore the special property of the optimal solution of Problem \ref{p:I_lower_convex}, with a given $\tau_t$, we investigate the following optimization problem
\begin{prob}\label{p:I_lower_lower}

\begin{subequations}
	\begin{align}
		E_{I}^{\text{loc}}\left( {\tau_t, n} \right)  \triangleq &\mathop {\min }\limits_{\{f_i\}} \sum\limits_{i = 1}^{n-1} {k_0 f_i^2{l_i}} \nonumber \\
		\text{s.t.} \quad &0 \leq {f_i} \le {f_{\max}}, \forall i \in \{1, 2, ..., n-1\}, \label{e:l_lower_lower_box}\\
		&\sum\limits_{i=1}^{n-1}{\frac{l_i}{f_i}} \le
		T_{\text{th}}- \tau_t - \sum\limits_{i = n}^N {\frac{l_i}{f_e}}  \label{e:I_lower_lower_f_cons}.
	\end{align}
	\end{subequations}
	
\end{prob}

In Problem \ref{p:I_lower_convex}, since $f_i \leq f_{\max}$ for $i \in \{1, 2, ..., n-1\}$ according to (\ref{e:I_lower_convex_f_interval}), $\tau_t$ has to be no larger than $\left(T_{\text{th}} - \sum_{i=1}^{n-1}\frac{l_i}{f_{\max}} - \sum_{i=n}^{N} \frac{l_i}{f_e}\right)$ according to (\ref{e:I_lower_convex_active_inequality}).
With this implicit constraint on $\tau_t$ and the definition of $E_I^{\text{loc}}(\tau_t, n)$ in Problem \ref{p:I_lower_lower}, Problem \ref{p:I_lower_convex} is equivalent with the following optimization problem
\begin{prob} \label{p:I_lower_upper}

\begin{subequations}
\begin{align}
E_{I} \left( n \right) = \mathop{\min} \limits_{\tau_t} & \quad \tau_t \left(e^{\frac{d_n}{W \tau_t}} - 1\right) \frac{1}{h} + E_{I}^{\text{loc}}\left( {\tau_t, n} \right)   \nonumber \\
	\text{s.t.} & \quad 0 \leq \tau_t \leq  T_{\text{th}} - \sum_{i=1}^{n-1}\frac{l_i}{f_{\max}} - \sum_{i=n}^{N} \frac{l_i}{f_e}.
	\end{align}
\end{subequations}

\end{prob}

Problem \ref{p:I_lower_lower} is convex and satisfies the Slater's condition, then the KKT condition can serve as a sufficient and necessary condition for the optimal solution
\cite{boyd_vandenberghe_2004}. Specifically, the Lagrange function of Problem \ref{p:I_lower_lower} can be written as

\begin{equation}
\begin{array}{ll}
 &L(\{f_i\}, \lambda, \{\mu_i\}, \{\nu_i\}) \\
=& \sum_{i=1}^{n-1} k_0 l_i f_i^2 - \sum_{i=1}^{n-1} \mu_i f_i  \\
+ & \lambda \left(\sum\limits_{i = 1}^{n - 1} \frac{l_i}{f_i} + \tau_t + \sum\limits_{i = n}^N {\frac{l_i}{f_e}}  - T_{\text{th}}\right) + \sum_{i=1}^{n-1}  \nu_i \left(f_i  - f_{\max}\right)
\end{array}
\end{equation}
where $\lambda \geq 0$ is the Lagrange multiplier associated with constraint (\ref{e:I_lower_lower_f_cons}), $\mu_i \geq 0$ and $\nu_i\geq 0$ are the Lagrange multipliers associated with constraints $f_i \geq 0$ and $f_i \leq f_{\max}$ in (\ref{e:l_lower_lower_box}) for $i \in \{1, 2, ..., n-1\}$, respectively.
Then the KKT condition can be listed as in (\ref{e:KKT})

\begin{subequations} \label{e:KKT}
\begin{align}
2k_0 f_i l_i - \lambda \frac{l_i}{f_i^2} - \mu_i + \nu_i = 0, \forall i \in \{1, 2, ..., n-1\} \label{e:KKT_Lagrange}\\
\sum\limits_{i = 1}^{n - 1} \frac{l_i}{f_i} + \tau_t + \sum\limits_{i = n}^N {\frac{l_i}{f_e}} \le T_{\text{th}}  \label{e:KKT_sum_cons}\\
0 \leq {f_i} \le {f_{\max}} , \forall i \in \{1,2,...,n-1\}  \label{e:KKT_box_cons}\\
\lambda \left(\sum\limits_{i = 1}^{n - 1} \frac{l_i}{f_i} + \tau_t + \sum\limits_{i = n}^N {\frac{l_i}{f_e}}  - T_{\text{th}}\right)  = 0 \label{e:KKT_slack_sum_cons}\\
\mu_i f_i, \forall i \in \{1, 2, ..., n-1\} \label{e:KKT_slack_lower_bound}\\
\nu_i \left(f_i  - f_{\max}\right), \forall i \in \{1, 2, ..., n\}  \label{e:KKT_slack_upper_bound}\\
\lambda \geq 0  \label{e:KKT_lambda}\\
\nu_i \geq 0, \mu_i \geq 0, \forall i \in \{1, 2, ..., n-1\} \label{e:KKT_mu_nu}
\end{align}
\end{subequations}

With KKT condition in (\ref{e:KKT}), the optimal solution of $f_i$ for $i\in \{1, 2, ..., n-1\}$ can be derived, which is given in the following theorem.
\begin{thm} \label{lem:I_optimal_f}
The optimal solution of Problem \ref{p:I_lower_lower} is $f_i = \frac{\sum_{i=1}^{n-1} l_i}{\left(T_{\text{th}} - \sum_{i=n}^{N} \frac{l_i}{f_e} - \tau_t\right)}$ for $i\in \{1, 2, ..., n-1\}$ when $\tau_t \in \left(0, T_{\text{th}} - \sum_{i=1}^{n-1} \frac{l_i}{f_{\max}} - \sum_{i=n}^{N}\frac{l_i}{f_e}\right]$.
\end{thm}
\begin{IEEEproof}
Please refer to Appendix \ref{app:l_optimal_f}.
\end{IEEEproof}

To this end, the optimal solution of Problem \ref{p:I_lower_lower} has been found, and its minimal achievable cost function can be written as
$E_I^{\text{loc}}(\tau_t, n) = {k_0 \left(\sum_{i=1}^{n-1} l_i \right)^3}/{\left(T_{\text{th}} - \sum_{i=n}^{N} {l_i}/{f_e} - \tau_t\right)^2}$
%
%
which is actually a convex function with $\tau_t$. Next we go back to solve Problem \ref{p:I_lower_upper}.
Given the expression of $E_I^{\text{loc}}(\tau_t, n)$, it can be easily seen that for $\tau \in [0, T_{\text{th}} - \sum_{i=1}^{n-1}\frac{l_i}{f_{\max}} - \sum_{i=n}^{N} \frac{l_i}{f_e}]$, the objective function of Problem \ref{p:I_lower_upper} is a convex function. Hence Problem \ref{p:I_lower_upper} is a one-dimensional convex optimization problem, whose optimal solution can be found by the Golden search method \cite{press2007numerical}. To this end, Problem \ref{p:I_lower_upper}, which is also equivalent to Problem \ref{p:I_lower}, has been solved optimally.
Notice that our presented method only requires one dimension numerical search for the optimal $\tau_t$ while the optimal solutions of all the other variables, i.e., $f_i$ for $i\in \{1, 2, ..., n-1\}$, are returned in closed-form. This procedure only involves a computation complexity of $O\left(\log\left(\frac{T_{\text{th}} - \sum_{i=1}^{n-1}\frac{l_i}{f_{\max}} - \sum_{i=n}^{N} \frac{l_i}{f_e}}{\delta}\right)\right)$, where $\delta$ is the tolerance. Compared with traditional numerical methods,
which have a computation complexity $O(n^3)$ \cite{boyd_vandenberghe_2004}, our presented method can largely reduce the computation complexity.

In terms of solving Problem \ref{p:I_upper}, by evaluating $E_I(n)$ for $n \in \mathcal{N}$, the $n$ that associates the minimal $E_I(n)$ can be recognized as the optimal solution of Problem \ref{p:I_upper}, and the optimal solution of Problem \ref{s:p:constant_gain} as well. At the end of this subsection, the whole procedure for solving Problem \ref{s:p:constant_gain} is summarized in the following algorithm.
\begin{algorithm}[H]
	\caption{The procedure for solving Problem \ref{s:p:constant_gain}.}
	\begin{algorithmic}[1]\label{a:I}
	 \FOR {$n = 1:N$}
	 \STATE {With the given $n$, use the Golden section method to find the optimal $\tau_t$ for Problem \ref{p:I_lower}, which is denoted as $\tau_t^*(n)$, and record the associated minimal achievable cost function $E_I(n)$.}
	 \ENDFOR
	 \STATE {Find $n^* = \mathop{\arg\min} \limits_{n \in \mathcal{N}} E_I(n)$.}
	 \STATE {Output $n^*$ as the optimal solution of $n$ for Problem \ref{s:p:constant_gain}, $\left(e^{\frac{d_{n^*}}{W \tau_t^*(n^*)}}-1\right)\frac{1}{h}$ as the optimal $P_t$ of Problem \ref{s:p:constant_gain}, $\frac{\sum_{i=1}^{n-1} l_i}{\left(T_{\text{th}} - \sum_{i=n}^{N} \frac{l_i}{f_e} - \tau_t^*(n^*)\right)}$  as the optimal $f_i$ of Problem \ref{s:p:constant_gain} for $i = 1, 2, ..., n^*-1$.}
	\end{algorithmic}
\end{algorithm}

\begin{remark} \label{r:mark_1}
In Algorithm \ref{a:I}, the computation complexity to evaluate $E_I(n)$ can be written as $O\left(\log\left(\left({T_{\text{th}} - \sum_{i=1}^{n-1}{l_i}/{f_{\max}} - \sum_{i=n}^{N} {l_i}/{f_e}}\right)/{\delta}\right)\right)$. Hence the whole computation complexity to find the optimal solution of Problem \ref{s:p:constant_gain} is $O\left(\sum_{n=1}^{N} \log\left(\left({T_{\text{th}} - \sum_{i=1}^{n-1} {l_i}/{f_{\max}} - \sum_{i=n}^{N} {l_i}/{f_e}}\right)/{\delta}\right)\right)$.
\end{remark}

\subsection{Optimal Solution of the Task Offloading and Resource Allocation Problem in Fast Fading Channels} \label{OS3}

In this subsection, the problem of task offloading and resource allocation in fast fading channels, which corresponds to Problem \ref{s:p:multiple}, would be solved.
In Problem \ref{s:p:multiple}, {$n\left(\bm{h}_1 \rightarrow h_{n, 1} \right)$}, $M(n)$, and $\tilde{t}_n$ are required to be solved but entangled in (\ref{e:s:p:multiple_deadline}).
{
To free the entanglement of these online policies and find their optimal solutions, we decompose Problem \ref{s:p:multiple} into two levels. In the lower level, under any realization of $\bm{h}_1$, $\bm{h}_2$, ..., $\bm{h}_N$, with $n\left(\bm{h}_1 \rightarrow h_{n, 1} \right)$ given, the rest variables, including $M(n)$ and $\tilde{t}_n$, are optimized.
In the upper level, $n\left(\bm{h}_1 \rightarrow h_{n, 1} \right)$ is optimized over any realization of $\bm{h}_1$, $\bm{h}_2$, ..., $\bm{h}_N$.
For the ease of presentation and without loss of correctness, $n$ and $n(\bm{h}_1\rightarrow h_{n,1})$ are used interchangeably in the following.}

{In the lower level, under any realization of $\bm{h}_1$, $\bm{h}_2$, ..., $\bm{h}_N$, with $n\left(\bm{h}_1 \rightarrow h_{n, 1} \right)$ given, Problem \ref{s:p:multiple} dwells into Problem \ref{o:p3:1} since only $Q_{n, 1}(d_n,h_{n,1}; M(n),\tilde{t}_n)$ involves $M(n)$ and $\tilde{t}_n$ in the objective function of Problem \ref{s:p:multiple}. The Problem \ref{o:p3:1} is given as follows}
\begin{prob}\label{o:p3:1}
\begin{subequations}
	\begin{align}
		Q(n,h_{n,1}) \triangleq \mathop {\min }\limits_{M(n),\tilde{t}_n} \quad & Q_{n, 1}(d_n,h_{n,1}; M(n),\tilde{t}_n)  \nonumber \\
\text{s.t.} \quad 		    &	T_3(n, M(n), \tilde{t}_n) \leq T_{\text{th}},  \label{e:Q_n_T_3}\\
		           & M(n) \in \mathcal{I}, \\
                   & 0 < \tilde{t}_n  \leq \tau,  \label{e:Q_n_t}
	\end{align}
	\end{subequations}
\end{prob}
{
where the newly defined $Q(n, h_{n,1})$ represents the energy consumption for offloading sub-task $\phi_n$'s computing data with the measurement of $h_{n,1}$.
\footnote{

{
According to the definition of $Q(n, h_{n,1})$ in Problem \ref{o:p3:1}, it seems that $M(n)$ and $\tilde{t}_n$ only depend on $h_{n,1}$ when $n(\bm{h}_1\rightarrow h_{n,1})$ is given. It should be noticed that this does not mean $M(n)$ and $\tilde{t}_n$ is irrelevant to $\bm{h}_1$, $\bm{h}_2$, ..., $\bm{h}_{n-1}$. In essence, $\bm{h}_1$, $\bm{h}_2$, ..., $\bm{h}_{n-1}$ indirectly determine $M(n)$ and $\tilde{t}_n$, which is given in such a way: They first affect $n(\bm{h}_1\rightarrow h_{n,1})$ and then the $n(\bm{h}_1\rightarrow h_{n,1})$ affects $M(n)$ and $\tilde{t}_n$. Hence the definition of $Q(n, h_{n,1})$ given in Problem \ref{o:p3:1} is not contradict with the claim after Problem \ref{s:p:multiple} and $M(n)$ and $\tilde{t}_n$ are still the policies depending on $\bm{h}_1$, $\bm{h}_2$, ..., ${h}_{n,1}$.
}
}
}

{
In the upper level, $n\left(\bm{h}_1 \rightarrow h_{n, 1} \right)$ is required to be optimized to minimize the objective function of Problem \ref{s:p:multiple} over any realization of $\bm{h}_1$, $\bm{h}_2$, ..., $\bm{h}_N$, which can be also written as $\left(Q(n, h_{n,1}) + \sum\limits_{i = 1}^{n- 1} {{l_i}  f_l^2 k_0}  \right)$ since $Q(n, h_{n,1})$ represents the minimized $Q_{n, 1}(d_n,h_{n,1}; M(n),\tilde{t}_n)$ over $M(n)$ and $\tilde{t}_n$. In such a case, it is equivalent to solve Problem \ref{s:p:multiple_two_level}, which is given as follows}
\begin{prob}\label{s:p:multiple_two_level}
\begin{subequations}
	\begin{align}
		\mathop {\min }\limits_{n} \quad & \mathbb{E}_{{\bm{h}_1, \bm{h}_2, ..., \bm{h}_N}} \left\{Q(n, h_{n,1}) + \sum\limits_{i = 1}^{n- 1} {{l_i}  f_l^2 k_0}  \right\} \nonumber \\
		\text{s.t.} \quad & {n(\bm{h}_1\rightarrow h_{n,1}) \in \mathcal{N}}.
	\end{align}
\end{subequations}
\end{prob}

We start by solving Problem \ref{o:p3:1}. For Problem \ref{o:p3:1}, to find its optimal solution, the following lemmas can be expected.

\begin{lem}\label{lem:OS3:1-1}
With $n$, $m$, $h$, and $d$ given, where $n\in \mathcal{N}$, $m \in \{1, 2, ..., M(n)\}$, and $d \geq 0$, there is $Q_{n,m}(d,h; M(n), t_1) \geq Q_{n,m}(d,h; M(n), t_2)$ for $0 \leq t_1 \leq t_2 \leq \tau$.
\end{lem}
\begin{IEEEproof}
Please refer to Appendix \ref{app:OS3:1-1}.
\end{IEEEproof}

\begin{lem}\label{lem:OS3:1-2}
With $n$ and $d$ given, where $n\in \mathcal{N}$ and $d\geq 0$,
for $m_1 \leq m_2$,
there is $Q_{n,1}(d,h; m_1, \tau) \geq Q_{n,1}(d,h; m_2, \tau)$.
\end{lem}
\begin{IEEEproof}
Please refer to Appendix \ref{app:OS3:1-2}.
\end{IEEEproof}

Lemma \ref{lem:OS3:1-1} implies that the function $Q_{n,m}(d,h; M(n), t)$ is a non-increasing function with $t$ for $t \in [0, \tau]$. Note that the original domain of definition for $t$ is $(0, \tau]$, while the case $t=0$ is considered in Lemma \ref{lem:OS3:1-1}. It should be noticed that $Q_{n,m}(d,h;M(n),t) = Q_{n,m}(d,h; M(n)-1, \tau)$ when $t=0$. Hence the case that $t=0$ is not weird. What is even further, the case that $t=0$ can facilitate the proof for Theorem \ref{lem:OS3:thm}.
Lemma \ref{lem:OS3:1-2} shows that the function $Q_{n,1}(d,h; m', \tau)$ is a non-increasing function with $m'$.
With the help of Lemma \ref{lem:OS3:1-1} and Lemma \ref{lem:OS3:1-2}, Theorem \ref{lem:OS3:thm} can be expected, which can help to generate the optimal solution of Problem \ref{o:p3:1}.
\begin{thm} \label{lem:OS3:thm}
To solve Problem \ref{o:p3:1}, it is optimal to select $M(n)\in \mathcal{I}$ and $\tilde{t}_n \in (0, \tau]$ such that
$T_3(n, M(n), \tilde{t}_n) = T_{\text{th}}$.
\end{thm}
\begin{IEEEproof}
Please refer to Appendix \ref{app:OS3:thm}.
\end{IEEEproof}

With Theorem \ref{lem:OS3:thm}, the optimal $M(n)$ and $t$ of Problem \ref{o:p3:1} have been found for a given $n$, which is defined as $M^*(n)$ and $\tilde{t}_n^*$ respectively. In the next step, we need to solve Problem \ref{s:p:multiple_two_level}. For Problem \ref{s:p:multiple_two_level}, the method of dynamic programming can be utilized to find the optimal solution \cite{10.5555/526593}.

Specifically,
define

\begin{equation}  \label{e:Z_end}
Z_N(h_{N,1}) = \min\left(Q_{N,1}(d_N, h_{N,1}; M^*(N), \tilde{t}_N^*), k_0 l_n f_l^2\right),
\end{equation}

\begin{equation} \label{e:Z_n_int}
\begin{split}
& Z_n(h_{n,1}) \\
=&  \min \left(Q_{n,1}(d_n, h_{n,1}; M^*(n), \tilde{t}_n^*), k_0 l_n f_l^2 + Z_{n+1}\right), \\
& \quad \quad \quad \quad \quad \quad \quad \quad  \quad \quad \quad \quad \quad \quad \quad \quad \forall n \in \mathcal{N} \setminus \{N\}.
\end{split}
\end{equation}

and

\begin{equation} \label{e:Z_n_iter}
Z_n = \int_{0}^{\infty} Z_n(h_{n,1}) p(h_{n,1}) d h_{n,1}, \forall n \in \mathcal{N}.
\end{equation}

It can be speculated that $Z_n(h_{n,1})$ is the minimal expected forthcoming energy consumption of the IoT device when the IoT device has just completed the local computing of sub-task $\phi_{n-1}$ and has measured the channel gain of $h_{n,1}$ for $n \in \mathcal{N}$.
In addition, $Z_n$ is the minimal expected forthcoming energy consumption of the IoT device when the IoT device has completed the local computing of sub-task $\phi_{n-1}$ but has no knowledge of $h_{n,1}$ for $n\in \mathcal{N}$.
For the ease of presentation, we also define $Z_{N+1}=0$. Then (\ref{e:Z_n_int}) will also be effective for $n=N$.

To achieve minimal energy consumption, the IoT device should decide whether to offload the input data of the next sub-task after it completes local computing of every sub-task, until it has made the decision to stop local computing. Suppose the IoT device has just completed the local computing of sub-task $\phi_{n-1}$ and has measured the channel gain $h_{n,1}$, then it will stop local computing and starts task offloading of sub-task $\phi_{n}$ if $Q_{n,1}(d_n, h_{n,1}; M^*(n), \tilde{t}_n^*) < \left(k_0 l_n f_l^2 + Z_{n+1}\right)$ and continue local computing otherwise.
In summary, the solving procedure of Problem \ref{s:p:multiple} can be summarized in Algorithm \ref{a:3} as follows.

\begin{algorithm}[H]
	\caption{The procedure for solving Problem \ref{s:p:multiple}.}
	\begin{algorithmic}[1]\label{a:3}
	  \STATE{Offline Part:}
		\STATE{Utilizing (\ref{e:Q_int}), (\ref{e:Q_iterative}), (\ref{e:Q_init}) and with the aid of Theorem \ref{lem:OS3:thm} to calculate $Q_{n,m}(d_n, h_{n,1}; M(n)^*, \tilde{t}_n^*)$ for $n \in \mathcal{N}$ and $m \in \{1, 2, ..., M^*(n)\}$. Mark $Q_{n,1}(d_n, h_{n,1}; M(n)^*, \tilde{t}_n^*)$ as $Q_{n,1}(d_n, h_{n,1})$ for $n\in \mathcal{N}$. Utilizing (\ref{e:Z_end}), (\ref{e:Z_n_int}) and (\ref{e:Z_n_iter}) to calculate $Z_{n}(h_{n,1})$ and $Z_n$ for $n \in \mathcal{N}$.
Store the calculated $Z_n$, and $Z_n(h_{n,1})$, $Q_{n,m}(d_n)$, $Q_{n,m}(d_n, h_{n,1})$ and the calculated optimal $d_{n,m}$ for solving the optimization problem in (\ref{e:Q_iterative}) at the IoT device before it starts computing the whole task.
Store the calculated $Z_n$, and $Z_n(h_{n,1})$ for $n \in \mathcal{N}$ and $m\in \{1, 2, ..., M^*(n)\}$ at the IoT device.}	
		\STATE{Online Part:}
		\WHILE{$n \leq N$ and the IoT device has not stopped local computing}
		\IF{$ Q_{n,1}(d_n, h_{n,1}; M^*(n), \tilde{t}_n^*) < \left(k_0 l_n f_l^2 + Z_{n+1}\right)$}
		\STATE{Stop local computing and upload sub-task $\phi_n$ to the edge server. In the procedure of task offloading, adopt the stored optimal solution of $d_{n,m}$ for solving (\ref{e:Q_iterative}) to realize the expected energy consumption $Q_{n,1}(d_n, h_{n,1}; M^*(n), \tilde{t}_n^*)$. }
		\ELSE
		\STATE{Complete the computing of sub-task $\phi_{n}$ at local.}
		\ENDIF
		\STATE{$ n = n + 1$}
		\ENDWHILE
	\end{algorithmic}
\end{algorithm}

\begin{remark} \label{r:mark_2}
{
In terms of computation complexity of Algorithm \ref{a:3}, exponential complexity would be involved in the procedure of iterative computing for $Q_{n,m}(d,h; m, t)$. However, this part of computation belongs to offline computing, which can be completed by the edge server and does not impose a computational burden on the IoT device. With a table of $Q_{n,m}(d,h; m, t)$ stored at the IoT device, it only needs to make a comparison between two values in each step, whose computation complexity is ignorable.}
\end{remark}

\subsection{Further Discussion for Fast Fading Channels} \label{s:further_discussion}
In fast fading channels, the $\tau$ may be very small compared with the time scale for offloading the input data of one sub-task. To get some insight into this case, we will analyze the trend when $\tau$ approaches zero in the following. Suppose the IoT device selects to offload the data for computing sub-task $\phi_n$. As $\tau$ converges to zero, $M(n)$ will converge to infinity and the $\tilde{t}_n$ can be taken as $\tau$ without loss of generality. In this case, looking into the expression of $Q_{n,1}(d, h_{n,1}; M(n), \tau)$, there is

\begin{equation}  \label{e:Q_n_1_extreme}
\begin{split}
& Q_{n,1}(d, h_{n,1}; M(n), \tau) = \\
&
{
\mathop{\min} \limits_{0 \leq d_{n,1} \leq d} \left(e(d_{n,1}, h_{n,1}, \tau) + Q_{n,2}(d- d_{n,1}; M(n), \tau) \right).
}
\end{split}
\end{equation}

As $\tau$ approaches zero, the optimal $d_{n,1}$ for achieving $Q_{n,1}(d, h_{n,1}; M(n), \tilde{t}_n)$, denoted as $d^*_{n,1}$, would also converge to zero. Otherwise an infinite amount of data would be offloaded within time $T(n)$. Hence there is $Q_{n,2}(d- d^*_{n,1}; M(n), \tilde{t}_n)=Q_{n,2}(d; M(n), \tilde{t}_n)$.
Combing with the fact that $Q_{n,1}(d, h_{n,1}; M(n), \tau) \leq Q_{n,2}(d; M(n), \tau)$ according to (\ref{e:Q_com_m}) and $e(d^*_{n,1}, h_{n,1}, t_{n,1})\geq 0$. It can be easily derived that $e(d^*_{n,1}, h_{n,1}, t_{n,1}) = 0$.
Therefore, there is

\begin{equation} \label{e:Q_equi_inf}
\begin{split}
  Q_{n,1}(d, h_{n,1}; M(n), \tilde{t}_n) = & Q_{n,2}(d; M(n), \tilde{t}_n) \\
= & Q_{n,1}(d; M(n), \tilde{t}_n).
\end{split}
\end{equation}

The equation (\ref{e:Q_equi_inf}) indicates that there is no difference in the expected energy consumption for task offloading  no matter how much the $h_{n,1}$ is.
In this case, we only need to evaluate the minimal expected energy consumption, i.e., $Q_{n,1}(d; M(n), \tilde{t}_n)$, rather than measure the $h_{n,1}$, before deciding to offload the sub-task $\phi_n$.
However, since $M(n)$ is approaching infinity, it is not practical to calculate $Q_{n,1}(d; M(n), \tilde{t}_n)$ by the iterative method presented in (\ref{e:Q_iterative}).
To overcome this problem, recalling the fact that there is an infinite number of independently faded blocks within time $T(n)$, it is reasonable to assume that the process of $h_{n,1}, h_{n,2}, ..., h_{n, M(n)}$ are subject to an ergodic process.
Therefore, within time $T(n)$, the amount of data for task offloading can be written as $\mathbb{E}\{W\log\left(1 + p_n(h)\right)\}T(n) $, where $p_n(h)$ represents IoT device transmit power when instant channel state is $h$, and the expected energy consumption can be written as $\mathbb{E}\{p_n(h)\} T(n)$.
Then the energy consumption minimization problem for offloading $d_n$ nats can be reformulated as follows
\begin{prob}\label{s:p:ergodic1}

\begin{subequations}
\begin{align}
	\mathop{\min} \limits_{\{p_n(h)| h \in \mathcal{H}\}} & \quad  \mathbb{E} \left\{p_n(h)  \right\} T(n)  \nonumber \\
	\text{s.t.} & \quad
	 \mathbb{E} \left\{W\log \left(1 + p_n(h)\right)\right\} T(n) \geq d_n, \\
     & \quad p_n(h) \geq 0, \forall h \in \mathcal{H},
\end{align}
\end{subequations}

\end{prob}
which is equivalent with the following optimization problem
\begin{prob}\label{s:p:ergodic2}

\begin{subequations}
\begin{align}
	\mathop{\min} \limits_{\{p_n(h)| h \in \mathcal{H}\}} & \quad  \mathbb{E} \left\{p_n(h)  \right\} \nonumber \\
	\text{s.t.} & \quad
	 \mathbb{E} \left\{ \log \left(1 + p_n(h)\right)\right\} \geq \frac{d_n}{T(n)W}, \label{s:p:ergodic2_offload_data}\\
	& \quad p_n(h) \geq 0, \forall h \in \mathcal{H}.
\end{align}
\end{subequations}

\end{prob}

Problem \ref{s:p:ergodic2} is a convex optimization problem and falls into the category of the stochastic optimization problem.
The Lagrangian of Problem \ref{s:p:ergodic2} can be written as
%
$\mathcal{L}\left(\{p_n(h)\},  \zeta \right) = \mathbb{E} \{p_n(h) \}- \zeta \left(\mathbb{E} \{ \log\left(1 + p_n(h) h \right)\} - \frac{d_n}{T(n) W}\right)$,
%
where $\zeta$ is the non-negative Lagrange multiplier associated with the constraint (\ref{s:p:ergodic2_offload_data}).
Then the Lagrange dual function can be written as

\begin{equation}
D(\zeta) = \mathop {\min} \limits_{\{p_n(h) \geq 0 | h \in \mathcal{H}\}}  \mathcal{L}\left(\{p_n(h)\},  \zeta \right).
\end{equation}

Given the fact that Problem \ref{s:p:ergodic2} is a convex optimization problem and the Slater condition is satisfied \cite{boyd_vandenberghe_2004}, the maximum of the function $D(\zeta)$ over $\zeta \geq 0$ is also the minimum of Problem \ref{s:p:ergodic2}.
To find the maximum of $D(\zeta)$ over $\zeta \geq 0$, a sub-gradient method can be utilized \cite{bertsekas2003convex}, which updates the $\zeta$ in the following way until the $\zeta$ converges

\begin{equation} \label{e:multiplier_update}
\zeta(i+1) = \left(\zeta(i) - a(i) \left( \frac{d_n}{T(n)W} - \mathbb{E} \{\log (1 + p_n(h,i) h )\} \right)\right)^+.
\end{equation}

In (\ref{e:multiplier_update}), $\zeta(i)$ is the Lagrange multiplier $\zeta$ at $i$th iteration,
$p_n(h,i)$ is the optimal $p_n(h)$ to achieve $D(\zeta(i))$ at $i$th iteration for $h\in \mathcal{H}$,
$a(t)$ is the positive step size at $i$th iteration,
and $(x)^+ = \max (x, 0)$. It should be noticed that $a(t)$ should satisfy $\sum_{i=0}^\infty a(i) = \infty$ and $\sum_{i=0}^{\infty} a(t)^2 < \infty$. One candidate $a(i)$ is $a(i)=1/i$.

For a given $\zeta$, to achieve $D(\zeta)$, the optimal $p_n(h)$ of the following optimization problem for every realization of $h$ should be derived
\begin{prob} \label{p:Lagrange_one_realization}

\begin{subequations}
\begin{align}
\mathop{\min} \limits_{p_n(h)} \quad &  p_n(h) - \zeta \log\left( 1 + p_n(h)\right) + \zeta \frac{d_n}{T(n) W} \nonumber \\
\text{s.t.} \quad & p_n(h) \geq 0.
\end{align}
\end{subequations}

\end{prob}

For Problem \ref{p:Lagrange_one_realization}, the optimal $p_n(h)$ can be easily obtained by setting the derivative of its objective function over $p_n(h)$ to be zero when $p_n(h) \geq 0$, which can be presented as
%
$p_n(h) = \left(\zeta - \frac{1}{h}\right)^+$.
%

To this end, Problem \ref{s:p:ergodic1} has been solved and how much energy would be consumed for offloading $d_n$ nats when the length of one fading block $\tau$ goes to zero has been answered.
Note that in this case, the minimal energy consumption for offloading $d_n$ nats is a fixed number, which can be denoted as $Q(n)$, rather than a random number depending on $h_{n,1}$ anymore.
Hence the optimal $n$ can be easily found among $\{1, 2, ..., N\}$ by selecting the $n$ with the minimal associated total energy consumption, which can be written as $\sum_{i=1}^n l_i f_l^2 k_0 +  Q(n)$.

\begin{remark} \label{r:mark_3}
According to the discussion in this subsection, it can be found that as $\tau$ decreases to zero, the online policy derived in Section \ref{OS3}, which depends on the instant $h_{n,1}$ in the $n$th step, trends to converge to an offline policy that can predetermine the optimal $n$ before the IoT device starts to run the whole computation task.
Compared with online policy, offline policy can save the calculation of $Q_{n,m}(d, h; M(n), \tilde{t}_n)$ for every $m\in \{1, ..., M(n)\}$ and $h\in \mathcal{H}$, which will involve a lot of computation.
Hence when $\tau$ is relatively small compared with the time scale for task offloading, offline policy discussed
in this subsection would be preferred.
\end{remark}

\section{Numerical Results} \label{s:numerical_results}

In this section, numerical results are presented. Default system parameters are set as follows.
By referring to the setup in \cite{6846368}, a computation task with 10 sub-tasks is adopted.
The input data size $d_n$ of these 10 sub-tasks are given as $\{36, 22, 30, 6, 47, 30, 5, 47, 14, 49\}$ k bits, and the workload ${l_n}$ for $n\in \mathcal{N}$ are selected to be
$\{7, 30, 25, 16, 32, 15, 37, 44, 24, 40\}$ M CPU cycles.
Similar to \cite{7442079}, $k_0$ is set as $10^{-28}$.
The channel coherence time, i.e., the length of one fading block, $\tau$ is set as 20ms.
The deadline for completing the whole computation task $T_{\text{th}}$ is selected to be 350ms.
\footnote{This time delay is necessary by evaluating the computation capability of the edge server and the workload of the investigated computation task.}.
The CPU frequency of the edge server $f_e=3$ GHz.
The maximal computation capability of the IoT device $f_{\max}=500$ MHz.
The computation capability of the IoT device in the case of fast fading channels $f_l$ is also set as 500 MHz \cite{6846368}.
The bandwidth $W=1$ MHz.
In the case of fast fading channels, the channel is Rayleigh distributed, which means that the channel gain $h_{n,m}$ is subjected to an exponential distribution.
The mean of $h_{n,m}$ is set as 50 \footnote{{Considering the fact that $h_{n,m}$ actually represents the received SNR when the transmit power of the IoT device is 1, the selection of this number level is reasonable.}}.
{All the simulations are run on a desktop with the CPU of Intel i7-7700K working at a frequency of 4.2 GHz.}

\subsection{Slow Fading Channels} \label{s:num_slow}

\begin{figure}
	\centering
	\subfigure[Energy consumption versus $f_e$.]{
		\begin{minipage}[b]{0.5\textwidth}
			\centering
			\includegraphics[width=0.8\figwidth]{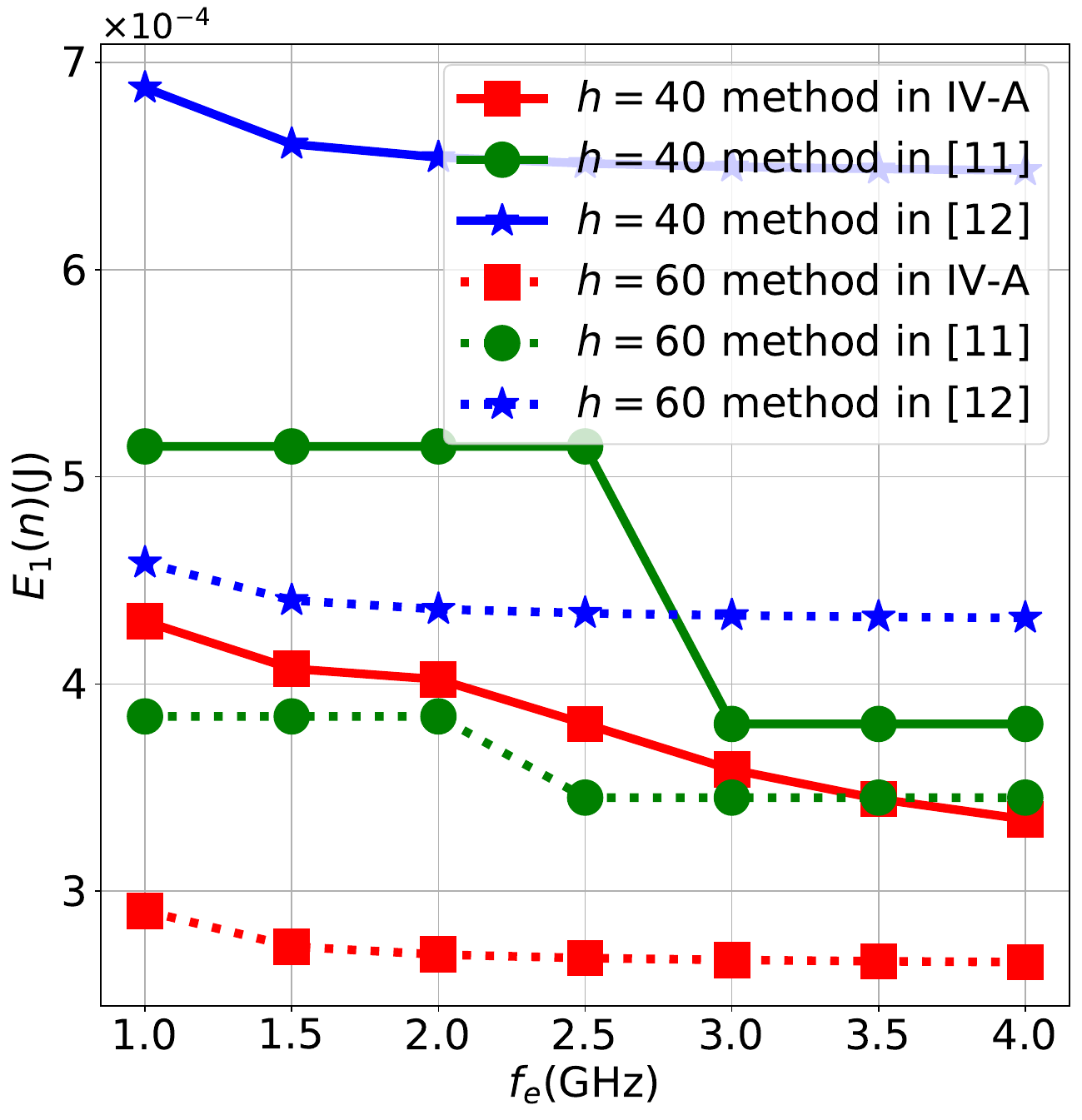}
			\label{NR1-1}
		\end{minipage}	
		}				
	\subfigure[Energy consumption versus $T_{\text{th}}$.]{
		\begin{minipage}[b]{0.5\textwidth}
			\centering
			\includegraphics[width=0.8\figwidth]{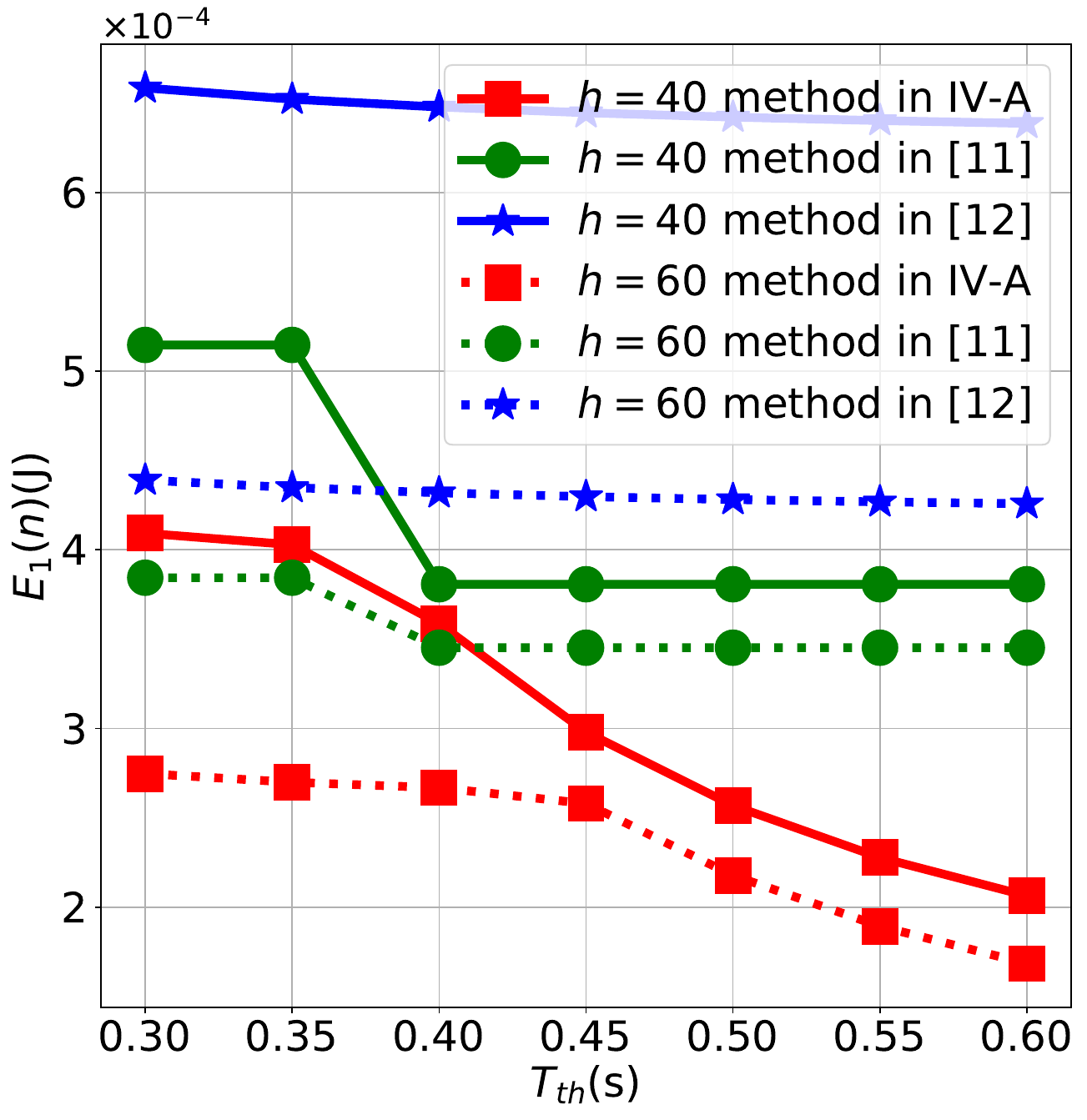}
			\label{NR1-2}
		\end{minipage}
			}
	\subfigure[Running time vs. Numbers of sub-tasks $N$.]{
		\begin{minipage}[b]{0.5\textwidth}
			\centering
			\includegraphics[width=0.8\figwidth]{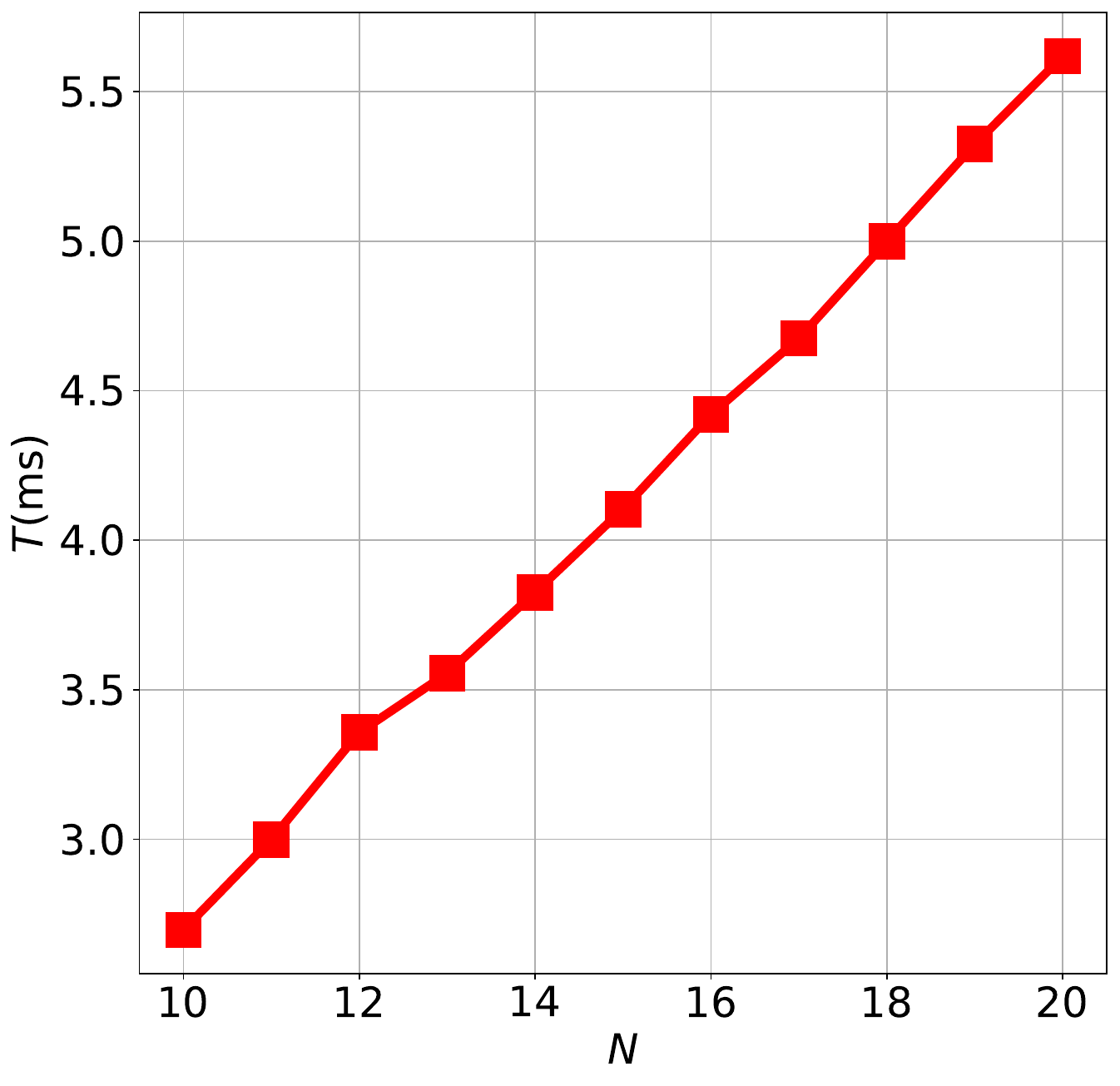}
			\label{NR1-3}
		\end{minipage}
		}
	\caption{Performance analysis under slow fading channel.}
	\label{f:performance_slow_fading}
	\end{figure}

{Fig. \ref{NR1-1} compares the IoT device's energy consumption between our proposed method in Section \ref{OS1} and the methods in \cite{6846368}
 and \cite{6574874}, which represent existing strategies dealing with sequential task and binary offloading respectively, under various edge server's CPU frequency $f_e$ \footnote{{The other literature dealing with the sequential task, including \cite{6849257} and \cite{8854339}, are not compared due to similarity with the method in \cite{6846368} or model mismatch. }}.
For the method in \cite{6846368}, the transmit power, CPU frequency at the IoT device and the edge server are not optimized, so we adopt its default setting\cite{6846368}.
For the method in \cite{6574874}, we utilize its energy consumption model to generate an offloading decision in the framework of binary offloading.
It can be observed that our proposed method always outperforms the methods in \cite{6846368} and \cite{6574874}.
This result can be also found in Figs. \ref{NR1-2}, \ref{NR2-1}, and \ref{NR2-2}.}
These results prove the effectiveness of our proposed method.
Additionally, it can be seen that
the energy consumption under the method in \cite{6846368} goes down like a ``staircase''.
This can be explained as follows: The transmit power and CPU frequency are fixed in \cite{6846368}. Hence the energy consumption can be only reduced when the IoT device finds a better occasion to offload (by deciding to offload since another sub-task) for the method in \cite{6846368}.
Similar trends can be also found in Fig. \ref{NR1-2} due to the same reason, which will not be discussed anymore in the part of Fig. \ref{NR1-2}.
It can be also seen that as the CPU frequency of the edge server $f_e$ grows, the associated energy consumption trends to decrease for the proposed method in Section \ref{OS1}.
This is because the increase of $f_e$ can contribute to the enlargement of Problem \ref{s:p:constant_gain}'s feasible region.
Hence the energy consumption under our proposed method, which is also the minimum of Problem \ref{s:p:constant_gain}, would decrease.

{
Fig. \ref{NR1-2} plots IoT device's energy consumption with our proposed method in Section \ref{OS1} and the methods in \cite{6846368} {and \cite{6574874}} versus the deadline for completing the whole computation task $T_{\text{th}}$, when the channel gain $h$ is set as 40 and 60, respectively.}
From Fig. \ref{NR1-2}, we can observe that for a fixed channel gain, the IoT device's minimal energy consumption decreases as $T_{\text{th}}$ grows. This is due to the fact that the increase of $T_{\text{th}}$ can help to relax the feasible region of Problem \ref{s:p:constant_gain}. Hence the IoT device's minimal energy consumption, which is also the minimum of Problem \ref{s:p:constant_gain}, would decrease.

{
Fig. \ref{NR1-3} plots the running time when the number of sub-tasks $N$ varies. When $N>10$, we rerun the default computation task. It can be seen that the average running time to work out Problem \ref{s:p:constant_gain} is around 3.0 ms, which is ignorable compared with the scale of $T_{\text{th}}$.
This is attributed to the fact that one Golden search, which is simple to realize,  is enough to find out the optimal solution of Problem \ref{s:p:constant_gain}, as shown in this paper. Thanks to the fast-growing manufacturing process of CPU in recent years, the computation capability of one IoT device nowadays does not lose to our desktop running the simulation. Hence we can claim the computation burden for working out the optimal solution of Problem \ref{s:p:constant_gain}
at the IoT device is ignorable when evaluating the time consumption for completing the computation task.
}


\subsection{Fast Fading Channels} \label{s:num_fast}


\begin{figure}
\centering
\subfigure[Energy consumption versus $f_e$.]{
	\begin{minipage}[b]{0.5\textwidth}
		\centering
		\label{NR2-1}
        \includegraphics[width=0.7\textwidth]{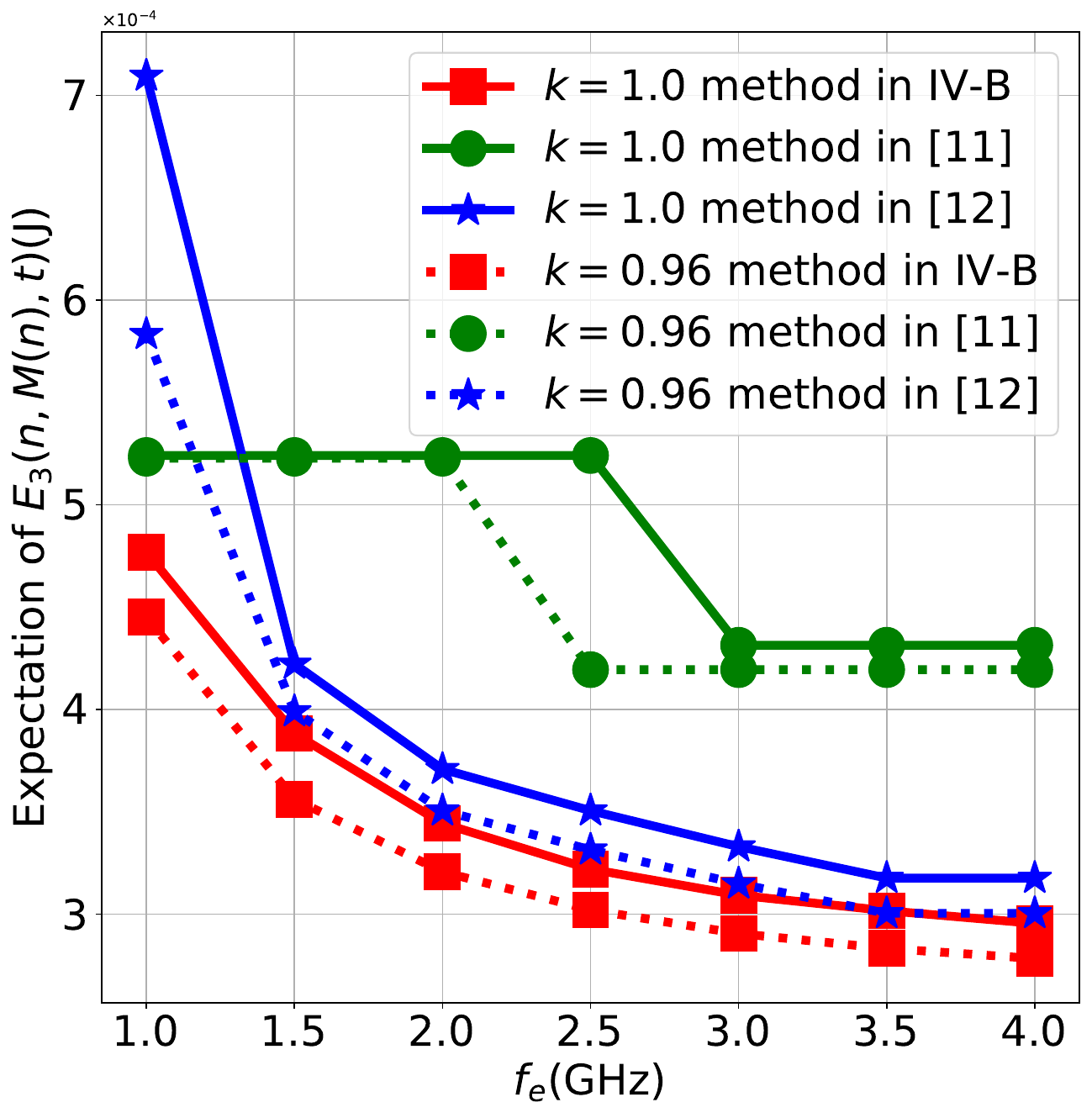}
	\end{minipage}
    }				
\subfigure[Energy consumption versus $T_{\text{th}}$.]{
	\begin{minipage}[b]{0.5\textwidth}
		\centering
		\label{NR2-2}
        \includegraphics[width=0.7\textwidth]{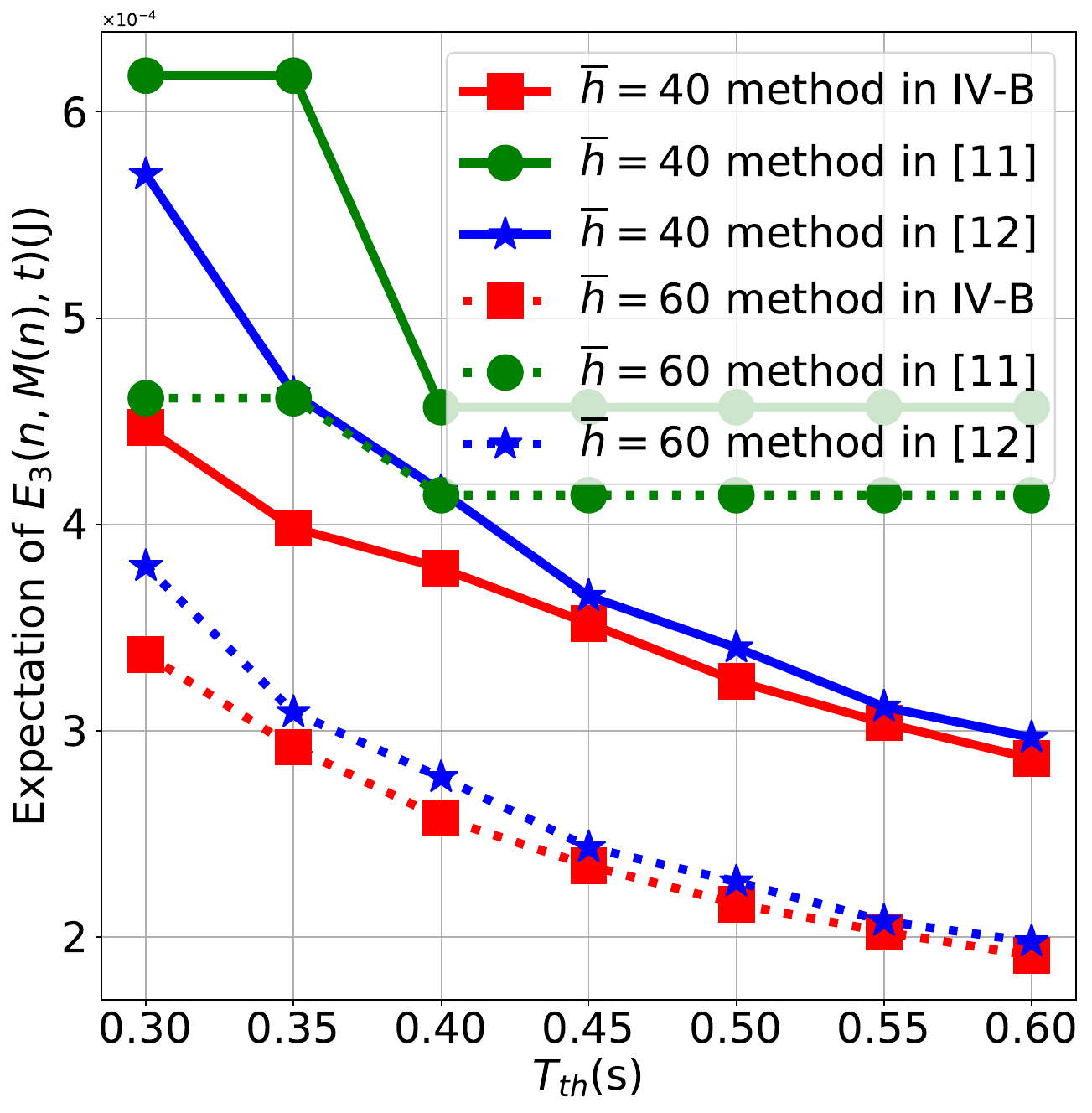}
	\end{minipage}
    }
\subfigure[Running time vs. Numbers of sub-tasks $N$.]{
	\begin{minipage}[b]{0.5\textwidth}
		\centering
		\label{NR2-3}
		\includegraphics[width=0.7\textwidth]{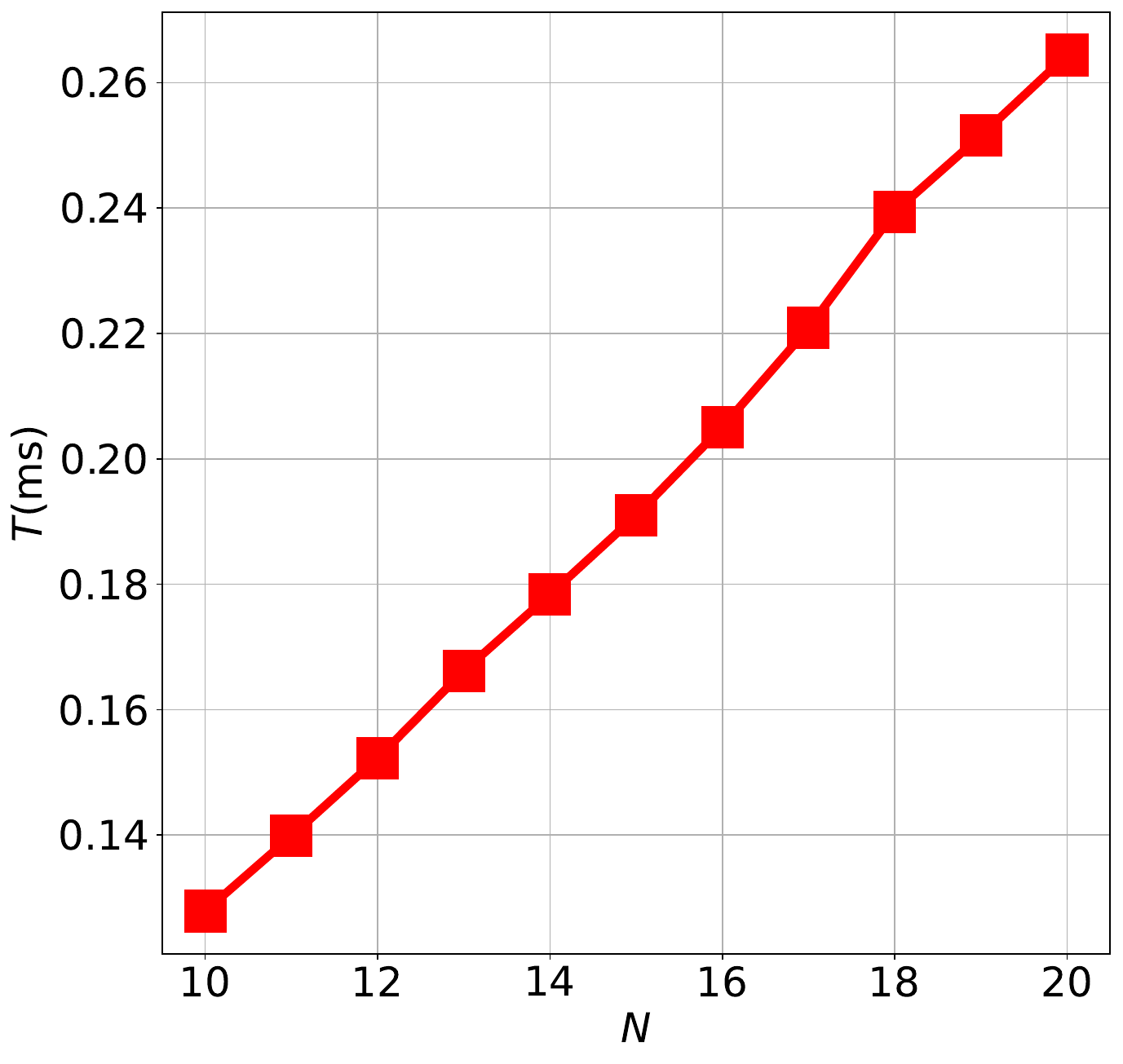}
	\end{minipage}
	}
\caption{Performance analysis under fast fading channel.}
\label{f:performance_fast_fading}
\end{figure}

{
Fig. \ref{NR2-1} illustrates IoT device's expected energy consumption {for our proposed method in Section \ref{OS3} and the methods in \cite{6846368} and \cite{6574874}} when edge server's CPU frequency $f_{e}$ varies.}
In the legend of Fig. \ref{NR2-1}, ''$k=1.0$" and"$k=0.96$'' indicate the scaling factor imposed on the input data size of every sub-task.
In terms of the trends of the methods in \cite{6846368} and \cite{6574874}, the energy consumption for the method in \cite{6846368} still goes down like a ``staircase'', which can be explained as the same way for Fig. \ref{NR1-1}, while the energy consumption for the method in \cite{6574874} gradually decrease with $f_e$.
The reason behind the behavior of the method in \cite{6574874} can be explained as follows.
In \cite{6574874}, a table is also built between expected energy consumption and data amount for offloading, which represents both continuity and convexity for the expected energy consumption with respect to the data amount for offloading as we have in our proposed method. Therefore a similar trend with our proposed method can be seen for the method in \cite{6574874}. On the other hand, the table building procedure does not consider the optimization of transmit power and will surely lead to higher energy consumption compared with our proposed method. Hence the method in \cite{6574874} will always have weaker performance than our proposed method. Similar trends can be also found in Fig. \ref{NR2-2} based on the same reason, which is not further discussed in the part of Fig. \ref{NR2-2}.
It can be also observed that IoT device's expected energy consumption decreases as $f_e$ grows, which can be explained in the same way as the case in the slow fading channel.

{Fig. \ref{NR2-2} plots IoT device's expected energy consumption for our proposed method in Section \ref{OS3} and the methods in \cite{6846368} and \cite{6574874} when $T_{\text{th}}$ varies. It can be observed that IoT device's expected energy consumption for our proposed method will decrease with $T_{\text{th}}$ when the mean of channel gain is set as 40 and 60, respectively.}
This can be explained as follows. Enlarged $T_{\text{th}}$ can help to relax the feasible region of
Problem \ref{s:p:multiple}. Hence the expected energy consumption, which is also the minimum of Problem \ref{s:p:multiple}, would decrease.

{
Fig. \ref{NR2-3} illustrates the average running time of online computing for solving Problem \ref{s:p:multiple} as the number of sub-tasks $N$ varies. When $N>10$, we rerun the default computation task. It can be observed that the running time is at the scale of 0.14 ms, while the offline computing time is at the scale of 1793.47 ms.
As discussed in Remark \ref{r:mark_2}, offline computing can be completed in advance by the edge server. Hence the computation burden for the IoT device, which merely comes from online computing, is ignorable (considering the rough equivalence of the computation capability between the IoT device and our simulating desktop as discussed in Section \ref{s:num_slow}) when evaluating the time consumption for solving Problem \ref{s:p:multiple}.
}
%

\begin{figure}
\centering
\subfigure[Verification of Lemma \ref{lem:OS3:1-1} and Lemma
	\ref{lem:OS3:1-2}.]{
		\begin{minipage}[b]{0.5\textwidth}
			\centering
			\label{NR3-1}
        	\includegraphics[width=0.7\textwidth]{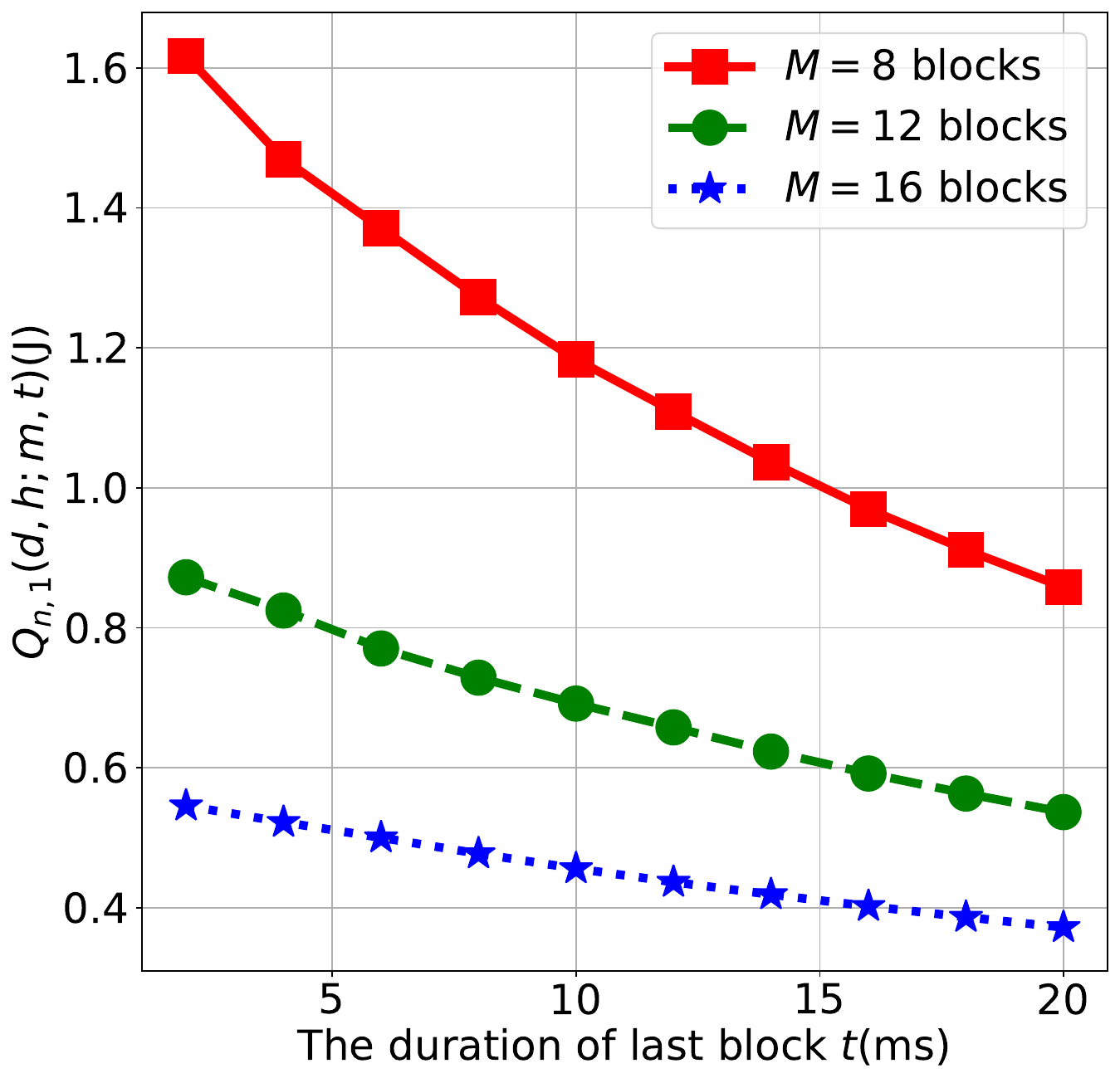}
		\end{minipage}
        }				
\subfigure[Verification of the discussion in \ref{s:further_discussion}.]{
	\begin{minipage}[b]{0.5\textwidth}
		\centering
		\includegraphics[width=0.7\textwidth]{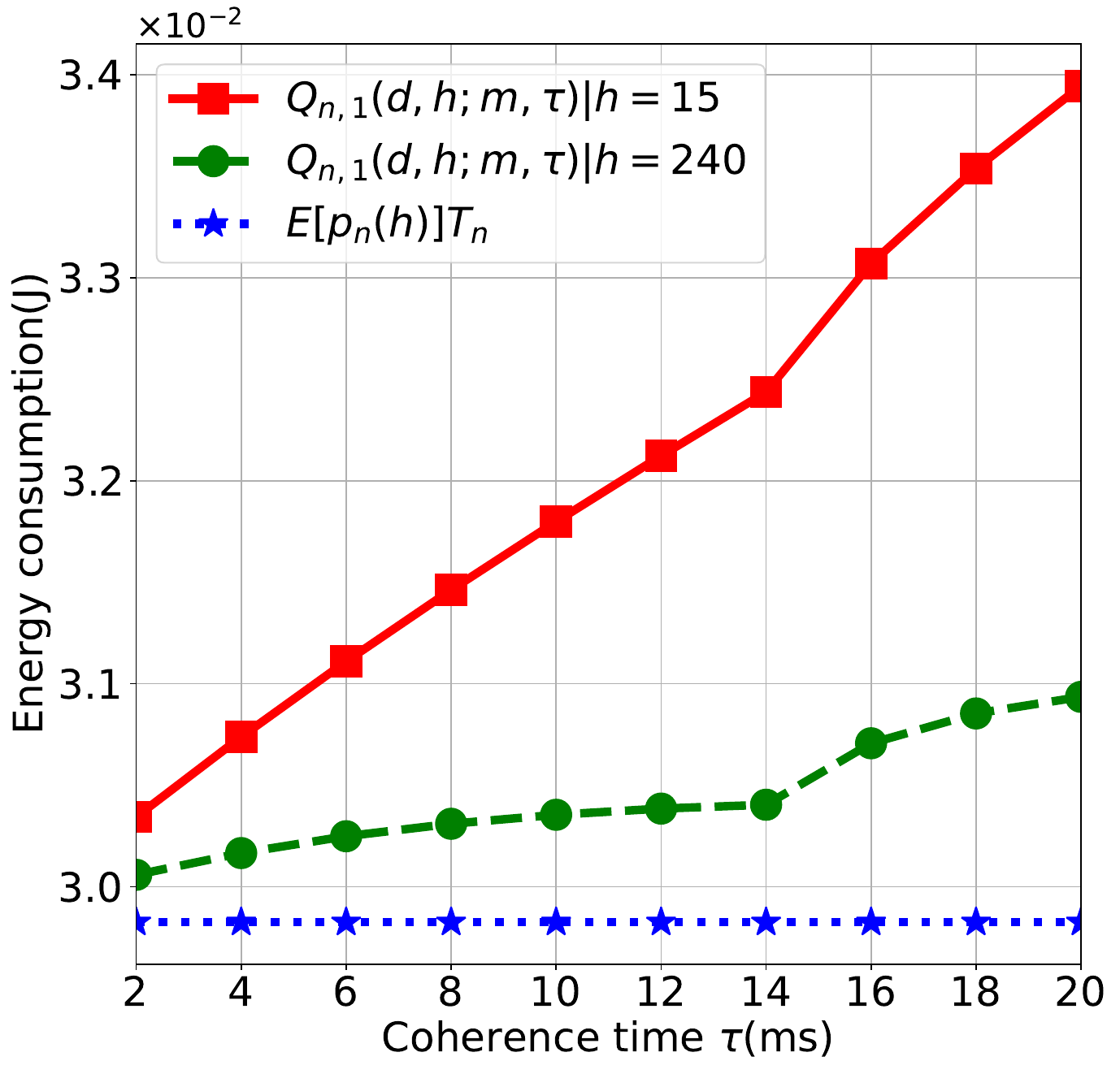}
		\label{NR3-2}
	\end{minipage}	
        }
\caption{Verification of lemmas and discussion.}
\label{f:verification}
\end{figure}

Fig. \ref{NR3-1} plots the function $Q_{n,1}(d,h; m, t)$ with $d = 1.2 \times 10^6$
bits when the duration of the last fading block $t$ varies from 5 ms to 20 ms.
The channel gain of the first fading block $h$ is set as 60.
It can be observed that the $Q_{n,1}(d,h; m, t)$ degrades as the duration of the last block $t$ increases when the number of fading blocks is fixed, which verifies the claim in Lemma \ref{lem:OS3:1-1}.
It can be also checked that the $Q_{n,1}(d,h; m, t)$ will decrease with the number of fading blocks when the duration of the last fading block $t$ is fixed, which verifies the claim in Lemma \ref{lem:OS3:1-2}.

{
In Fig. \ref{NR3-2}, the minimal expected energy consumption for offloading $d_n = 1.2 \times 10^6$
bits, i.e., \\
$Q_{n,1}(d_n, h_{n,1}; M(n), \tau)$,  under the various selection of $h_{n,1}$ is plotted versus channel coherence time $\tau$ when $n=1$. The $T(n)$ is set as $840$ms.}
As a comparison, the minimal $\mathbb{E}\{p_n(h) T(n)\}$ by solving Problem \ref{s:p:ergodic1} is also plotted. It can be observed that as $\tau$ decreases, $Q_{n,1}(d, h_{n,1}; M(n), \tau)$, no matter what $h_{n,1}$ is, trends to converge to the value of minimal $\mathbb{E}\{p_n(h) T(n)\}$. This verifies the discussion in Section \ref{s:further_discussion}.

\section{Conclusion} \label{s:conclusion}

In this paper, we considered a MEC system with sequentially dependent computation tasks under both slow and fast fading channels {with the purpose of supporting intelligent applications for an IoT network}. To minimize IoT device energy consumption, the optimization of task offloading strategy, communication resource allocation, and computation resource allocation are performed. For slow fading channels, with a task offloading decision given, a simple and optimal solution of CPU frequency for local computing and transmit power for task offloading were found by dividing the associated optimization problem into two levels. {The Golden search method is shown to be able to find the optimal solution in the upper level and a closed-form solution is derived in the lower level.}
Then the optimal task offloading decision can be easily found by one-dimensional search. For fast fading channels, an optimal online task offloading policy in response to the instant channel state was derived. It was also disclosed that the derived online policy trends to be an offline policy as the channel coherence time approaches zero, which can help to save the computation complexity when the channel coherence time is small enough. Our research results could provide helpful insights for
 {a MEC-supported IoT network running intelligent applications with sequential tasks.}

\begin{appendices}

\section{Proof of Theorem \ref{lem:I_optimal_f}} \label{app:l_optimal_f}
Since $f_i \in [0, f_{\max}]$ for $i \in \{1, 2, ..., n-1\}$, the discussion will be unfolded from three possible cases: 1) $f_i = 0$; 2) $0<f_i< f_{\max}$; 3) $f_i = f_{\max}$. For the first case, it is evident that the optimal $f_i$ for Problem \ref{p:I_lower_lower} cannot be zero for $i \in \{1, 2, ..., n-1\}$. Otherwise, the objective function of Problem \ref{p:I_lower_lower} would be infinite. Hence the case that $f_i=0$ for $i\in \{1, 2, ..., n-1\}$ can be precluded.
Define $\mathcal{A}_1(n-1) = \{i | 0<f_i< f_{\max}, i \in \{1, 2, ..., n-1\}\}$ and $\mathcal{A}_2(n-1) = \{i | f_i = f_{\max}, i \in \{1, 2, ..., n-1\}\}$, then there is $\mathcal{A}_1(n-1) \cup \mathcal{A}_2(n-1) = \{1, 2, ..., n-1\}$.

With regard to the second case, we can first derive that $\mu_i=\nu_i=0$ for $i \in \mathcal{A}_1(n-1)$ according to (\ref{e:KKT_slack_lower_bound}) and (\ref{e:KKT_slack_upper_bound}). Therefore, for $i\in \mathcal{A}_1(n-1)$, Eqn. (\ref{e:KKT_Lagrange}) dwells into
%
$2 k_0 f_i l_i = \lambda \frac{l_i}{f_i^2}, \forall i \in \mathcal{A}_1(n-1)$,
%
which further indicates that

\begin{equation} \label{e:I_lem_f_case_II_Lagrange}
f_i = \left(\frac{\lambda}{2k_0}\right)^{\frac{1}{3}}, \forall i \in \mathcal{A}_1(n-1).
\end{equation}

In other words, for $i \in \mathcal{A}_1(n-1)$, all the $f_i$s are equal.

With regard to the third case, it can be also derived that $\mu_i=0$ for $i\in \mathcal{A}_2(n-1)$ according to (\ref{e:KKT_slack_lower_bound}). Then Eqn. (\ref{e:KKT_Lagrange}) turns to
%
$2 k_0 f_{i} l_i = \lambda \frac{l_i}{f_{i}^2} - \nu_i, \forall i \in \mathcal{A}_2(n-1)$,
%
which can be further written as

\begin{equation} \label{e:I_lem_f_case_III_Lagrange}
\lambda = 2k_0 f_{\max}^3 + \frac{f_{\max}^2\nu_i}{l_i}, \forall i \in \mathcal{A}_2(n-1)
\end{equation}

since $f_i= f_{\max}$ for $i \in \mathcal{A}_2(n-1)$.
On the other hand, Eqn. (\ref{e:I_lem_f_case_II_Lagrange}) indicates that

\begin{equation} \label{e:I_lem_f_case_II_Lagrange_trans}
\lambda = 2 k_0 f_i^3, \forall i \in \mathcal{A}_1(n-1).
\end{equation}

Combing (\ref{e:I_lem_f_case_III_Lagrange}), (\ref{e:I_lem_f_case_II_Lagrange_trans}), and the facts that $f_i < f_{\max}$ for $i \in \mathcal{A}_1(n-1)$ and $\nu_j \geq 0$ for $j \in \mathcal{A}_2(n-1)$, there is
%
$\lambda = 2 k_0 f_i^3 < 2 k_0 f_{\max}^3 \leq 2k_0 f_{\max}^3 + \frac{f_{\max}^2\nu_j}{l_j} = \lambda$,
%
which contradicts the fact the $\lambda=\lambda$.
Hence the case that $\mathcal{A}_1(n-1)\neq \emptyset$ and $\mathcal{A}_2(n-1)\neq \emptyset$ cannot coexist.
In other words, only two possible cases for the optimal solution of Problem \ref{p:I_lower_lower} could happen: 1) $\mathcal{A}_1(n-1) = \{1, 2, ..., n-1\}$ while $\mathcal{A}_2(n-1) = \emptyset $ ; 2) $\mathcal{A}_2(n-1) = \{1, 2, ..., n-1\}$ while $\mathcal{A}_1(n-1) = \emptyset $. It should be noticed that all the $f_i$s are equal in every possible case.

On the other hand, we notice that the objective function of Problem \ref{p:I_lower_lower} is monotonically increasing function with $f_i$ while the left-hand side of constraint (\ref{e:I_lower_lower_f_cons}) is decreasing function with $f_i$ for $i\in \{1, 2, ..., n-1\}$. Hence to minimize the objective function of Problem \ref{p:I_lower_lower}, it would be optimal to decrease $f_i$ as much as possible, 
which will lead the activeness of constraint (\ref{e:I_lower_lower_f_cons}), i.e., the holding the following equality

\begin{equation} \label{e:I_f_cons_equality}
\sum\limits_{i=1}^{n-1}{\frac{l_i}{f_i}}  =
		T_{\text{th}}- \tau_t - \sum\limits_{i = n}^N {\frac{l_i}{f_e}}.
\end{equation}

As a summary, when $\tau_t = T_{\text{th}} - \sum_{i=1}^{n-1} \frac{l_i}{f_{\max}} - \sum_{i=n}^{N}\frac{l_i}{f_e}$, (\ref{e:I_f_cons_equality}) and the fact that $f_i$ are all equal can lead to the solution $f_i = \frac{\sum_{i=1}^{n-1} l_i}{\left(T_{\text{th}} - \sum_{i=n}^{N} \frac{l_i}{f_e} - \tau_t\right)} = f_{\max}$ for $i\in \{1, 2, ..., n-1\}$, which corresponds to the case that $\mathcal{A}_1(n-1) = \{1, 2, ..., n-1\}$ and $\mathcal{A}_2(n-1) = \emptyset$; when $0 < \tau_t < T_{\text{th}} - \sum_{i=1}^{n-1} \frac{l_i}{f_{\max}} - \sum_{i=n}^{N}\frac{l_i}{f_e}$, (\ref{e:I_f_cons_equality}) and the fact that $f_i$ are all equal can lead to the solution
$f_i = \frac{\sum_{i=1}^{n-1} l_i}{\left(T_{\text{th}} - \sum_{i=n}^{N} \frac{l_i}{f_e} - \tau_t\right)}$ for $i\in \{1, 2, ..., n-1\}$, which corresponds to the case that $\mathcal{A}_2(n-1) = \{1, 2, ..., n-1\}$ and $\mathcal{A}_1(n-1) = \emptyset$.


\section{Proof of Lemma \ref{lem:OS3:1-1}} \label{app:OS3:1-1}
We complete this proof via induction method.
In the first step, the case that $m= M(n)$ is investigated.
When $m=M(n)$, there is one fading block left for task offloading. In this case, there is $Q_{n,M(n)}(d,h; M(n), t) = e(d, h, t)$.
For the function $e(d, h, t)$, the first-order derivative of $e(d, h, t)$ with $t$ can be given as

\begin{equation} \label{e:diff_e_t}
\frac{d e(d, h, t)}{dt} = \frac{1}{h}\left(e^{\frac{d}{t}}-1-\frac{d}{t}\times
	e^{\frac{d}{t}}\right)
\end{equation}

Define $x \triangleq d/t \geq 0$, the term $\left(e^{\frac{d}{t}}-1-\frac{d}{t}\times
	e^{\frac{d}{t}}\right)$ in the right-hand side of (\ref{e:diff_e_t}) can be written as
$g(x) = e^x - 1 - xe^x$, whose first-order derivative $g'(x) = -xe^x \leq 0$. Then there is
$g(x) \leq g(0)  = 0 $ for $x\geq 0$, which proves that $\frac{de(d,h,t)}{dt} \leq 0$.
Hence for $0 \leq t_1 \leq t_2 \leq \tau$, there would be
$e(d, h, t_1) \geq e(d, h, t_2)$.

In the second step, we need to prove the holding of $Q_{n,m}(d, h; M(n), t_1) \geq Q_{n,m}(d, h; M(n), t_2)$ given the condition that $Q_{n, m+1}(d, h; M(n), t_1) \geq Q_{n,m+1}(d, h; M(n), t_2)$ for $0 \leq t_1 \leq t_2 \leq \tau$ and any $d$ and $h$ value.
Denote $d_1^*$ and $d_2^*$ as the optimal $d_{n,m}$ to achieve $Q_{n, m}(d, h; M(n), t_1)$ and
$Q_{n, m}(d, h; M(n), t_2)$ respectively, where the definition of $d_{n,m}$ can be found in (\ref{e:Q_iterative}), then there is

\begin{equation} \label{e:lem3-1-e1}
\begin{split}
	 & Q_{n, m}(d, h; M(n), t_1)\\
	=& e(d_1^*, h, \tau) + Q_{n, m+1}(d-d_1^*; M(n), t_1)\\
\geq & e(d_1^*, h, \tau)  + Q_{n, m+1}(d-d_1^*; M(n), t_2)  \\
\geq & e(d_2^*, h, \tau) + Q_{n,m+1}(d-d_2^*; M(n), t_2 ) \\
   = & Q_{n, m}(d, h; M(n), t_2).
\end{split}
\end{equation}
The first inequality of (\ref{e:lem3-1-e1}) comes from the fact that $Q_{n,m+1}(d-d_1^*; M(n), t_1) \geq Q_{n,m+1}(d - d_1^*; M(n), t_2)$ is a given condition.
The second inequality holds since $d_2^*$, rather than $d_1^*$, is the optimal solution of  $d_{n,m}$ to achieve $Q_{n, m}(d, h; M(n), t_2)$.


\section{Proof of Lemma \ref{lem:OS3:1-2}} \label{app:OS3:1-2}
We first prove that $Q_{n,1}(d; m-1, \tau) \geq Q_{n,1}(d; m, \tau)$.
Suppose $d^*_m$ is the optimal solution of $d_{n,1}$ to achieve $Q_{n,1}(d, h; m, \tau)$, where the definition of $d_{n,1}$ is given in (\ref{e:Q_iterative}). Then there would be

\begin{equation} \label{e:Q_com_m}
\begin{array}{ll}
    & Q_{n,1}(d, h; m, \tau) \\
  = & e(d^*_m, h, \tau) + Q_{n,2}(d - d^*_m; m, \tau) \\
 \overset{(a)} \leq & e(0, h, \tau) + Q_{n,2}(d; m, \tau)  \\
 =  & Q_{n,2}(d; m, \tau) \\
 =  & Q_{n,1}(d; m-1, \tau)
\end{array}
\end{equation}
{where $(a)$ comes from the fact that $d_m^*$ is the optimal solution of $d_m$ for minimizing $e(d_m, h, \tau) + Q_{n,2}(d-d_m; m, \tau)$ and it would achieve less cost for $d_m = d_m^*$ compared with the case with $d_m=0$.}

Hence according to (\ref{e:Q_com_m}), there is \footnote{
{
The inequality in (\ref{e:Q_com_m_final}) can be also envisioned intuitively: The function $Q_{n,1}(d; m, \tau)$ actually represents the minimum expected energy consumption for offloading $d$ nats over $(m-1)$ fading blocks, which would be surely no larger than $Q_{n,1}(d; m-1, \tau)$, which represents the minimum expected energy consumption for offloading the same amount of data over a shorter time interval.
}}

\begin{equation} \label{e:Q_com_m_final}
\begin{array}{ll}
     & Q_{n,1}(d; m, \tau) \\
   = & \int_{0}^{\infty} Q_{n,1}(d, h_{n,1}; m, \tau) p(h_{n,1}) dh_{n,1} \\
               \overset{(b)} {\leq} & \int_{0}^{\infty} Q_{n,1}(d; m-1, \tau) p(h_{n,1}) dh_{n,1} \\
               = & Q_{n,1}(d; m-1, \tau)
\end{array}
\end{equation}
{
where $(b)$ holds because of (\ref{e:Q_com_m}).
}

With the conclusion that $Q_{n,1}(d; m, \tau) \leq Q_{n,1}(d; m-1, \tau)$ derived in (\ref{e:Q_com_m_final}) and it can be further found that
\begin{equation} \label{e:Q_com_m_h}
\begin{array}{ll}
   & Q_{n,1}(d, h; m, \tau) \\
  = & e(d^*_m, h, \tau) + Q_{n,1}(d - d^*_m; m-1, \tau)\\
               \overset{(c)}{\leq} & e(d^*_{m-1}, h, \tau) + Q_{n,1}(d - d^*_{m-1}; m-1, \tau) \\
                \overset{(d)}{\leq} & e(d^*_{m-1}, h, \tau) + Q_{n,1}(d - d^*_{m-1}; m-2, \tau) \\
                = & Q_{n,1}(d, h; m-1, \tau)
\end{array}
\end{equation}
{
where the holding of (c) is due to a similar reason for (a) in (\ref{e:Q_com_m}), and the holding of (d) comes from (\ref{e:Q_com_m_final}).
}

Then with (\ref{e:Q_com_m_h}), it can be derived that for $m_1 \leq m_2$, there is

\begin{equation}
\begin{array}{ll}
   Q_{n,1}(d, h; m_1, \tau) & \geq Q_{n,1}(d, h; m_1+1, \tau) \\
  & \geq \cdots \\
  & \geq Q_{n,1}(d, h; m_2, \tau).
\end{array}
\end{equation}


\section{Proof of Theorem \ref{lem:OS3:thm}}  \label{app:OS3:thm}
Define $M^*(n)$ and $\tilde{t}_n^*$ as the  $M(n)$ and $\tilde{t}_n$ such that the constraint (\ref{e:Q_n_T_3}) is active, i.e., $T_3(n, M(n), \tilde{t}_n) = T_{\text{th}}$, respectively.
To prove this lemma, we need to prove that
\begin{equation}
Q_{n,1}(d_n, h; M^*(n), \tilde{t}_n^*) \leq Q_{n,1}(d_n, h; M'(n), \tilde{t}'_n)
\end{equation}
for any $M'(n) \in \mathcal{I}$ and $\tilde{t}'_n \in (0, \tau]$ such that
\begin{equation}
T_3(n, M'(n), \tilde{t}'_n) \leq T_{\text{th}}.
\end{equation}

Since $T_3(n, M^*(n), \tilde{t}_n^*) = T_{\text{th}}$ and $T_3(n, M'(n), \tilde{t}'_n) \leq T_{\text{th}}$, there is

\begin{equation} \label{e:M_compare}
\left(M^*(n) - 1\right) \tau + \tilde{t}_n^* \geq \left(M'(n) - 1\right) \tau + \tilde{t}'_n .
\end{equation}

Combing the fact that both $\tilde{t}_n^*$ and $\tilde{t}'_n$ falls into the interval $(0, \tau]$, then it can be concluded that
$M^*(n) \geq M'(n)$.
Then the discussion can be unfolded from the following two cases:
1). $M'(n) = M^*(n)$. In this case, there is $\tilde{t}_n^* > \tilde{t}'_n$ according to (\ref{e:M_compare}). Then according to Lemma \ref{lem:OS3:1-1}, there is $Q_{n,1}(d_n, h; M^*(n), \tilde{t}_n^*) \leq Q_{n,1}(d_n, h; M'(n), \tilde{t}'_n)$.
2). $M'(n) < M^*(n)$. In this case, it is hard to say $\tilde{t}_n^*> \tilde{t}'_n$ or $\tilde{t}_n^* \leq \tilde{t}'_n$, but there is $M'(n) \leq \left(M^*(n) - 1\right)$. Then there is
\begin{equation}
\begin{array}{ll}
  & Q_{n, 1}(d_n, h; M'(n), \tilde{t}'_n) \\
  \overset{(a)}{\geq} & Q_{n, 1}(d_n, h; M'(n), \tau) \\
  \overset{(b)}{\geq} & Q_{n,1}(d_n, h; M^*(n)-1, \tau) \\
  = &  Q_{n,1}(d_n, h; M^*(n), 0) \\
  \overset{(c)}{\geq}  & Q_{n,1}(d_n, h; M^*(n), \tilde{t}_n^*)
\end{array}
\end{equation}
{
where the holding of (a) and (c) can be explained by Lemma \ref{lem:OS3:1-1}, and (b) holds according to Lemma \ref{lem:OS3:1-2}.
}


\end{appendices}
\bibliographystyle{IEEEtran}

\bibliography{my.bib}

\begin{thebibliography}{10}
\providecommand{\url}[1]{#1}
\csname url@samestyle\endcsname
\providecommand{\newblock}{\relax}
\providecommand{\bibinfo}[2]{#2}
\providecommand{\BIBentrySTDinterwordspacing}{\spaceskip=0pt\relax}
\providecommand{\BIBentryALTinterwordstretchfactor}{4}
\providecommand{\BIBentryALTinterwordspacing}{\spaceskip=\fontdimen2\font plus
\BIBentryALTinterwordstretchfactor\fontdimen3\font minus
  \fontdimen4\font\relax}
\providecommand{\BIBforeignlanguage}[2]{{%
\expandafter\ifx\csname l@#1\endcsname\relax
\typeout{** WARNING: IEEEtran.bst: No hyphenation pattern has been}%
\typeout{** loaded for the language `#1'. Using the pattern for}%
\typeout{** the default language instead.}%
\else
\language=\csname l@#1\endcsname
\fi
#2}}
\providecommand{\BIBdecl}{\relax}
\BIBdecl

\bibitem{8391395}
P.~Porambage, J.~Okwuibe, M.~Liyanage, M.~Ylianttila, and T.~Taleb, ``Survey on
  multi-access edge computing for {Internet of Things} realization,''
  \emph{IEEE Commun. Surveys Tuts.}, vol.~20, pp. 2961--2991, June 2018.

\bibitem{zhou2019secure}
Z.~Zhou, B.~Wang, M.~Dong, and K.~Ota, ``Secure and efficient vehicle-to-grid
  energy trading in cyber physical systems: {Integration} of blockchain and
  edge computing,'' \emph{{IEEE} Trans. Syst., Man, Cybern., Syst.}, vol.~50,
  pp. 43--57, May 2020.

\bibitem{guo2022constructing}
T.~Guo, K.~Yu, M.~Aloqaily, and S.~Wan, ``Constructing a prior-dependent graph
  for data clustering and dimension reduction in the edge of {AIoT},''
  \emph{Future Generation Computer Systems}, vol. 128, pp. 381--394, Oct. 2021.

\bibitem{8016573}
Y.~{Mao}, C.~{You}, J.~{Zhang}, K.~{Huang}, and K.~B. {Letaief}, ``{A survey on
  mobile edge computing: The communication perspective},'' \emph{IEEE Commun.
  Surveys Tuts.}, vol.~19, pp. 2322--2358, Fourthquarter 2017.

\bibitem{feng2020attribute}
C.~Feng, K.~Yu, M.~Aloqaily, M.~Alazab, Z.~Lv, and S.~Mumtaz, ``Attribute-based
  encryption with parallel outsourced decryption for edge intelligent {IoV},''
  \emph{IEEE Trans. Veh. Technol.}, vol.~69, pp. 13\,784--13\,795, Nov. 2020.

\bibitem{7879258}
P.~{Mach} and Z.~{Becvar}, ``{Mobile edge computing: A survey on architecture
  and computation offloading},'' \emph{IEEE Commun. Surv. Tutor.}, vol.~19, pp.
  1628--1656, Thirdquarter 2017.

\bibitem{xu2021mcts}
J.~Xu, K.~Ota, M.~Dong, and H.~Zhou, ``{MCTS}-enhanced hybrid offloading for
  aerial multi-access edge computing,'' \emph{IEEE Wireless Commun.}, vol.~28,
  pp. 82--87, Oct. 2021.

\bibitem{wang2016mobile}
Y.~Wang, M.~Sheng, X.~Wang, L.~Wang, and J.~Li, ``{Mobile-edge computing:
  Partial computation offloading using dynamic voltage scaling},'' \emph{IEEE
  Trans. Commun.}, vol.~64, pp. 4268--4282, Oct. 2016.

\bibitem{8488502}
X.~{Cao}, F.~{Wang}, J.~{Xu}, R.~{Zhang}, and S.~{Cui}, ``Joint computation and
  communication cooperation for energy-efficient mobile edge computing,''
  \emph{IEEE Internet Things J.}, vol.~6, pp. 4188--4200, June 2019.

\bibitem{7842160}
Y.~{Mao}, J.~{Zhang}, S.~H. {Song}, and K.~B. {Letaief}, ``Power-delay tradeoff
  in multi-user mobile-edge computing systems,'' in \emph{2016 IEEE Global
  Communications Conference}, Washington, DC, USA, Dec. 2016, pp. 1--6.

\bibitem{8387798}
J.~{Ren}, G.~{Yu}, Y.~{Cai}, and Y.~{He}, ``Latency optimization for resource
  allocation in mobile-edge computation offloading,'' \emph{IEEE Trans.
  Wireless Commun.}, vol.~17, pp. 5506--5519, Aug. 2018.

\bibitem{7762913}
C.~{You}, K.~{Huang}, H.~{Chae}, and B.~{Kim}, ``Energy-efficient resource
  allocation for mobile-edge computation offloading,'' \emph{IEEE Trans.
  Wireless Commun.}, vol.~16, pp. 1397--1411, Mar. 2017.

\bibitem{mao2021computation}
S.~Mao, N.~Zhang, L.~Liu, J.~Wu, M.~Dong, K.~Ota, T.~Liu, and D.~Wu,
  ``Computation rate maximization for intelligent reflecting surface enhanced
  wireless powered mobile edge computing networks,'' \emph{IEEE Trans. Veh.
  Technol.}, vol.~70, pp. 10\,820--10\,831, Oct. 2021.

\bibitem{huang2020result}
W.~Huang, K.~Ota, M.~Dong, T.~Wang, S.~Zhang, and J.~Zhang, ``{Result return
  aware offloading scheme in vehicular edge networks for IoT},'' \emph{Computer
  Communications}, vol. 164, pp. 201--214, Oct. 2020.

\bibitem{ding2020deep}
F.~Ding, G.~Zhu, M.~Alazab, X.~Li, and K.~Yu, ``Deep-learning-empowered digital
  forensics for edge consumer electronics in {5G HetNets},'' \emph{{IEEE}
  Trans. Consum. Electron.}, early access, {Dec.}, {2020},
  doi:{10.1109/MCE.2020.3047606}.

\bibitem{SIG-039}
L.~Deng and D.~Yu, ``Deep learning: {Methods} and applications,''
  \emph{Foundations and Trends® in Signal Processing}, vol.~7, pp. 197--387,
  Jun. 2014.

\bibitem{6849257}
M.~{Jia}, J.~{Cao}, and L.~{Yang}, ``Heuristic offloading of concurrent tasks
  for computation-intensive applications in mobile cloud computing,'' in
  \emph{2014 IEEE Conference on Computer Communications Workshops (INFOCOM
  WKSHPS)}, Toronto, ON, Canada, May 2014, pp. 352--357.

\bibitem{6846368}
W.~{Zhang}, Y.~{Wen}, and D.~O. {Wu}, ``Collaborative task execution in mobile
  cloud computing under a stochastic wireless channel,'' \emph{IEEE Trans.
  Wireless Commun.}, vol.~14, pp. 81--93, Jan. 2015.

\bibitem{6574874}
W.~{Zhang}, Y.~{Wen}, K.~{Guan}, D.~{Kilper}, H.~{Luo}, and D.~O. {Wu},
  ``Energy-optimal mobile cloud computing under stochastic wireless channel,''
  \emph{IEEE Trans. Wireless Commun.}, vol.~12, pp. 4569--4581, Sept. 2013.

\bibitem{Add_NOMA_full_Ding}
Z.~{Ding}, P.~{Fan}, and H.~V. {Poor}, ``Impact of non-orthogonal multiple
  access on the offloading of mobile edge computing,'' \emph{IEEE Trans.
  Commun.}, vol.~67, pp. 375--390, Jan. 2019.

\bibitem{Add_NOMA_full_Peng}
B.~{Liu}, C.~{Liu}, and M.~{Peng}, ``Resource allocation for energy-efficient
  {MEC} in {NOMA}-enabled massive iot networks,'' \emph{IEEE J. Sel. Areas
  Commun.}, vol.~39, pp. 1015--1027, Apr. 2021.

\bibitem{Add_NOMA_partial_Wu}
Y.~{Wu}, K.~{Ni}, C.~{Zhang}, L.~P. {Qian}, and D.~H.~K. {Tsang},
  ``{NOMA}-assisted {Multi-Access} mobile edge computing: A joint optimization
  of computation offloading and time allocation,'' \emph{IEEE Trans. Veh.
  Technol.}, vol.~67, pp. 12\,244--12\,258, Dec. 2018.

\bibitem{8537962}
F.~{Wang}, J.~{Xu}, and Z.~{Ding}, ``Multi-antenna {NOMA} for computation
  offloading in multiuser mobile edge computing systems,'' \emph{IEEE Trans.
  Commun.}, vol.~67, pp. 2450--2463, Mar. 2019.

\bibitem{Add_energy_harvesting}
G.~{Zhang}, W.~{Zhang}, Y.~{Cao}, D.~{Li}, and L.~{Wang}, ``Energy-delay
  tradeoff for dynamic offloading in mobile-edge computing system with energy
  harvesting devices,'' \emph{IEEE Trans. Ind. Informat.}, vol.~14, pp.
  4642--4655, Oct. 2018.

\bibitem{gu2020mobile}
Q.~Gu, Y.~Jian, G.~Wang, R.~Fan, H.~Jiang, and Z.~Zhong, ``{Mobile edge
  computing via wireless power transfer over multiple fading blocks: An optimal
  stopping approach},'' \emph{IEEE Trans. Veh. Technol.}, vol.~69, pp.
  10\,348--10\,361, Sept. 2020.

\bibitem{8854339}
J.~{Yan}, S.~{Bi}, Y.~J. {Zhang}, and M.~{Tao}, ``Optimal task offloading and
  resource allocation in mobile-edge computing with inter-user task
  dependency,'' \emph{IEEE Trans. Wireless Commun.}, vol.~19, pp. 235--250,
  Jan. 2020.

\bibitem{BurdProcessor}
T.~D. Burd and R.~W. Brodersen, ``Processor design for portable systems,''
  \emph{J. VLSI Signal Process. Syst.}, vol.~13, pp. 203--221, Aug. 1996.

\bibitem{boyd_vandenberghe_2004}
S.~Boyd and L.~Vandenberghe, \emph{{Convex Optimization}}.\hskip 1em plus 0.5em
  minus 0.4em\relax Cambridge, England, UK: Cambridge University Press, 2004.

\bibitem{press2007numerical}
W.~Press, S.~Teukolsky, W.~Vetterling, and B.~Flannery, \emph{{Numerical
  Recipes 3rd Edition: The Art of Scientific Computing}}.\hskip 1em plus 0.5em
  minus 0.4em\relax Cambridge, England, UK: Cambridge University Press, 2007.

\bibitem{10.5555/526593}
D.~P. Bertsekas, \emph{{Dynamic Programming and Optimal Control}}.\hskip 1em
  plus 0.5em minus 0.4em\relax Nashua, NH, USA: Athena Scientific, 1995.

\bibitem{bertsekas2003convex}
D.~Bertsekas, A.~Nedi{\'c}, and A.~Ozdaglar, \emph{{Convex Analysis and
  Optimization}}.\hskip 1em plus 0.5em minus 0.4em\relax Nashua, NH, USA:
  Athena Scientific, 2003.

\bibitem{7442079}
C.~{You}, K.~{Huang}, and H.~{Chae}, ``Energy efficient mobile cloud computing
  powered by wireless energy transfer,'' \emph{IEEE J. Sel. Areas Commun.},
  vol.~34, pp. 1757--1771, May 2016.

\end{thebibliography}

\end{document}